\RequirePackage{fix-cm}
\documentclass[smallextended]{svjour3}       
\smartqed  
\usepackage{graphicx}
\usepackage{amsmath,amsfonts}
\begin{document}

\title{Mathematical models of Leukaemia and its treatment: A review
}


\author{S. Chuli\'an        \and   A. Mart\'inez-Rubio \and \\ M. Rosa \and    V.M. P\'erez-Garc\'ia  
}


\institute{S. Chuli\'an, A. Martínez-Rubio, M. Rosa \at
              Department of Mathematics, Universidad de C\'adiz, 11510 Puerto Real, C\'adiz, Spain \\
              \at
              Biomedical Research and Innovation Institute of C\'adiz (INiBICA), Hospital Universitario Puerta del Mar, 11009 C\'adiz, Spain
              \email{salvador.chulian@uca.es}           
           \and
          V.M. P\'erez-Garc\'{\i}a \at
          Instituto de Matem\'atica Aplicada a la Ciencia y la Ingenier\'ia (IMACI), Universidad de Castilla-La Mancha, 13005 Ciudad Real, Spain\\
          ETSI Industriales, Universidad de Castilla-La Mancha, 13005 Ciudad Real, Spain
}

\date{Received: date / Accepted: date}

\maketitle

\begin{abstract}
	Leukaemia accounts for around 3\% of all cancer types diagnosed in adults, and is the most common type of cancer in children of paediatric age. There is increasing interest in the use of mathematical models in oncology to draw inferences and make predictions, providing a complementary picture to experimental biomedical models. In this paper we recapitulate the state of the art of mathematical modelling of leukaemia growth dynamics, in time and response to treatment. We intend to describe the mathematical methodologies, the biological aspects taken into account in the modelling,  and the conclusions of each study. This review is intended to provide researchers in the field with solid background material, in order to achieve further breakthroughs in the promising field of mathematical biology.
\end{abstract}

\section{Introduction.}
Leukaemia is a cancerous disease in which blood cells display abnormal proliferation and invade other tissues. It is one of the biggest health issues globally. Almost half a million new leukaemia cases were diagnosed in 2018 \cite{Bray2018}.

Blood cancers affect the production and function of blood cells. They are the most common cancer types in children from birth to 14 years of age and account for around 3\% of all cancers diagnosed in developed countries. Blood cancer survival in adults is about 50\%. Although survival in children is higher and improving, blood cancer is still the major cause of cancer death in paediatric patients \cite{Miller2016,Desandes2016,Siegel2020}

Most types of blood cancer start in the bone marrow, which is where blood is produced. In most blood cancers, the normal development process, starting from stem cells and leading to a hierarchy of more differentiated cells, is interrupted by the uncontrolled abnormal growth of specific types of blood cell.

There are three major types of blood cancer. Leukaemias are caused by the rapid production of abnormal white blood cells. Lymphomas are a type of blood cancer comprising abnormal lymphocytes, a type of white blood cells that fight infections. These cells multiply and collect in lymph nodes and other tissues and impair the lymphatic system's functionality to remove unnecessary fluids from the body and fight infections. Finally, myeloma is a cancer of the plasma cells, which produce disease- and infection-fighting antibodies. 

\begin{figure}[ht]
	\centering\includegraphics[width=0.8\textwidth]{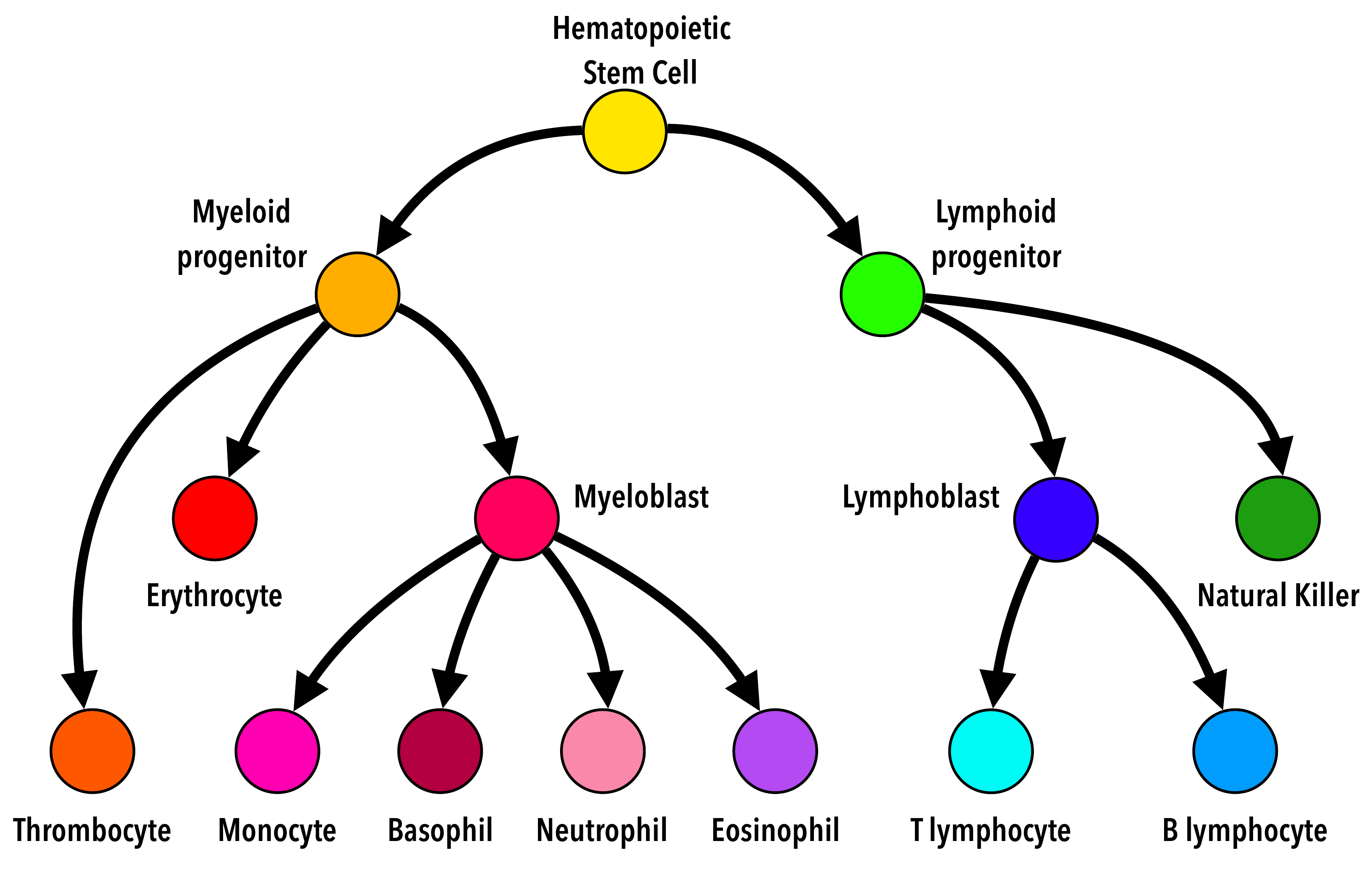}
	\caption{Differentiation tree for blood cells.}\label{Figure haematopoiesis}
\end{figure}

It is well understood how blood cells differentiate from stem cells into more specialised cells, as represented  schematically in Fig. \ref{Figure haematopoiesis}. At the top of the hierarchy governing normal haematopoiesis there are the haematopoietic stem cells (HSCs) \cite{Wang2005,Reya2001}. Pluripotent haematopoietic stem cells can give rise to either lymphoid or myeloid progenitors. Lymphoid progenitors can generate either lymphoblasts, which will become B or T lymphocytes, or Natural Killer cells. These are all part of the immune system. Myeloid progenitors can also lead to a broad variety of cells, including erythrocytes, thrombocytes, or other cells of the non-specific immune system.

\begin{figure}[ht]
	\centering 	\includegraphics[width=0.8\textwidth]{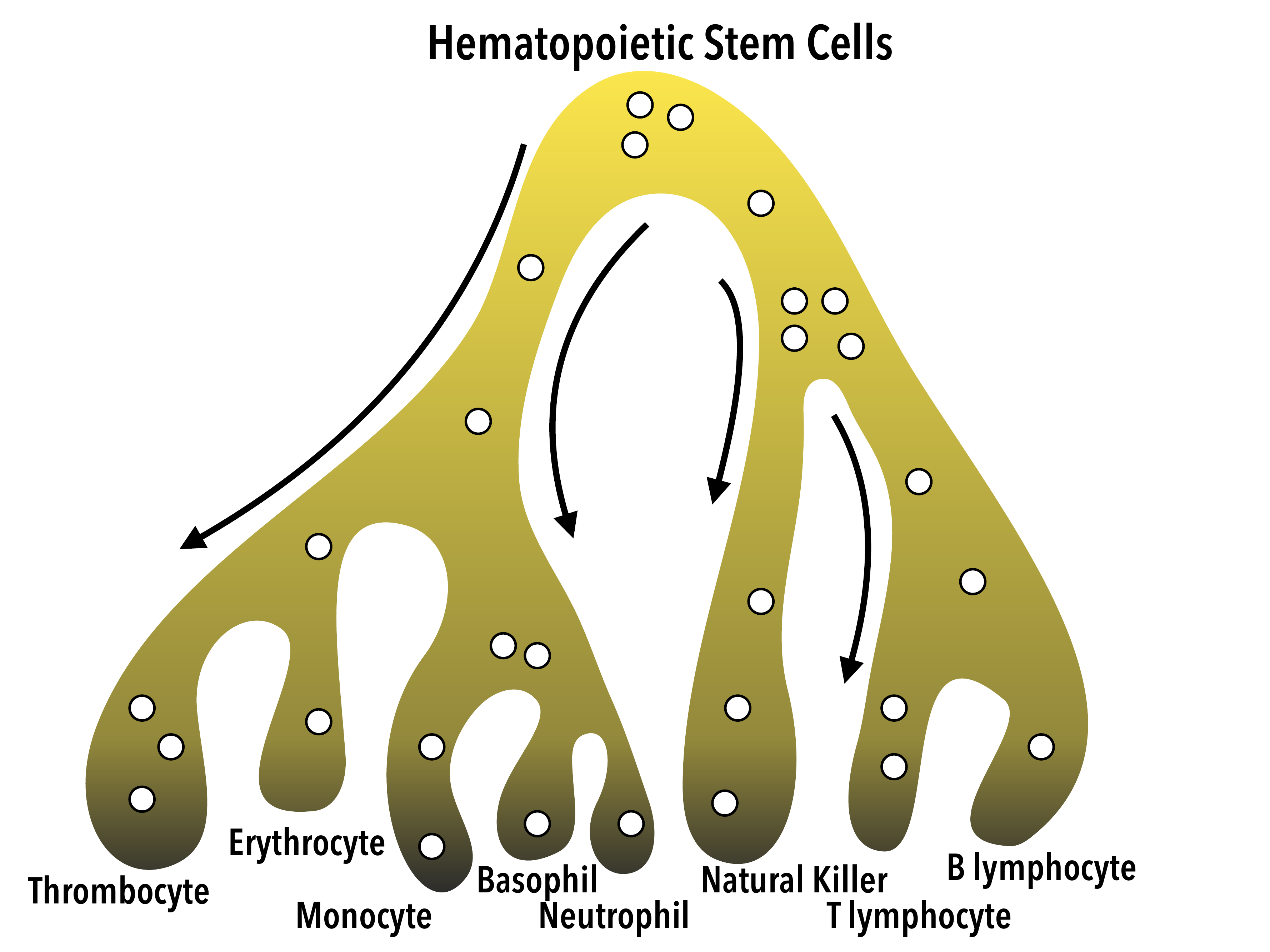} 
	\vspace{2pt}
	\caption{Representation of continuum differentiation model, where each dot represents a single cell and its location on a differentiation trajectory. Fig. adapted from Ref. \cite{Laurenti2018}.}\label{Figure Continuum}
\end{figure}

Although the classical understanding of haematopoiesis has considered cell types to be discrete compartments, current knowledge of the process \cite{Laurenti2018} considers the evolution of cell types as a continuum process (see Fig. \ref{Figure Continuum}). This is because haematopoietic cells acquire lineage features through a continuous process involving the expression of different characteristic molecules \cite{Velten2017}.

In this framework, the type of cell that becomes cancerous determines the specific type of blood cancer. For instance, leukaemia can be either myeloid (or myelogenous), or lymphoid (or lymphoblastic, or lymphocytic). Also, leukaemias can be distinguished by the maturation stage of the transformed cells. Acute leukaemias affect blast cells (immature blood cells), and grow very fast. 
Chronic leukaemias cause an accumulation of mature cells, leading to slowly growing cancers. Thus, there are four different classes of leukaemias: Acute Lymphoblastic leukaemia (ALL), Chronic Lymphocytic leukaemia (CLL), Acute Myelogenous leukaemia (AML) and Chronic Myelogenous leukaemia (CML) \cite{Fasano2017}. However, it is not completely clear whether the hierarchical organisation is preserved in blood cancers. Myeloid leukaemias seem to be hierarchically organised, whereas acute lymphoblastic leukaemias are not \cite{Rehe2012,Bonnet1997,Hope2004}.

The origins of the mutations are essential to understanding self-renewal and differentiation fractions for cancer cells \cite{Rehe2012,Passegue2003}.  These probabilities could be explained by some of the basic hallmarks of cancer \cite{Hanahan2011}, such as sustaining proliferative signalling, resisting cell death, immortality, evading growth suppressors, and metastasis. The detection of these hallmarks is essential to tailoring treatments, which depend on classifying each  patient within risk groups \cite{Cheok2006}.  Currently, patients are assigned to a  risk group depending on several factors, including the cell's morphology, the results of molecular or biochemical analysis, and the so-called flow cytometry techniques \cite{Maecker2012,Giner2002}. This is done by taking samples of the bone marrow (where the haematopoiesis process occurs) are obtained and characterised in terms of immunophenotypic patterns \cite{Lochem2004}, which can be standardised \cite{Dongen2012}. 

Mathematical modelling may offer a new perspective in Oncology, specifically in blood cancers, with a huge potential to develop new strategies to characterise tumours and personalise treatments \cite{Perez-Garcia2016,byrne2010dissecting,altrock2015mathematics}. The cancer hallmarks relevant to each specific stage of development for each tumour type can be accounted for in mathematical models, usually as parameters to be estimated or as equations which model the dynamics of blood cancer development.

Blood cancers, and specifically leukaemias, have been one of the first types of cancers that has been thoroughly studied by applied mathematicians. It should be noted that there are many mathematically-grounded studies published in this field in high impact medical and general-science journals. 

Leukaemias are a `global' disease of the bone marrow, and as such spatial effects are usually ignored. They can be modelled mathematically, in an initial approach, using ordinary differential equations. More complex models have used partial differential equations, but to describe the evolution of some kind of trait or subpopulation, rather than spatial variables. Furthermore, blood cell counts are an easy way to gather information about the evolution of the disease. Putting the data together has led to substantial interest in the disease from modellers and clinicians managing the disease.

Only a small fraction of the data available during routine clinical procedures is used for diagnosis, and incorporated into the models developed so far. This review focuses on the role of mathematical models based on  differential equations. However, mathematical techniques for (big-)data analysis  \cite{Saeys2016} also have huge potential for providing answers to specific questions of relevance to leukaemia, whether alone or in combination with other mathematical methods. For instance, some studies have pointed out their potential use in avoiding expert manual gating of the data to identify leukaemic clones \cite{Aghaeepour2013}, analysing mass cytometry data \cite{Weber2016}, or predicting treatment response \cite{deAndres-Galiana2015}. 

Our review is intended to expand the available literature on blood cancers \cite{Clapp2015,Fasano2017} to incorporate more studies and greater detail, by focusing on leukaemia. Our plan in this paper is as follows. Firstly, we summarise mathematical models based on differential equations describing the growth of myeloid leukaemias. This focus reflects the fact that myeloid leukaemias are the commonest among adults. The only models that exist for lymphoblastic leukaemia concern treatment. We then review mathematical models for different types of leukaemia treatment.  Finally, we discuss the results and summarise our conclusions.

\section{Mathematical models of myeloid leukaemias.}

Myeloid leukaemia arises from alterations of cells of the myeloid lineage, and is  considered a clonal disorder of the haematopoietic stem cells (HSCs). The condition may lead to an increase in  myeloid cell, erythroid cell or platelet counts, not only in peripheral blood but also in the bone marrow. As described above, the two general types are chronic myeloid leukaemia (CML) and acute myeloid leukaemia (AML), depending on the maturation stage of the cells. In CML cells mature during the chronic phase, while in AML blast cells fail to mature,  generating large amounts of blasts, i.e. immature cells \cite{Sawyers1999,Lowenberg1999}. 

\subsection{Stem-cell based models of myeloid leukaemias.}

Stem-cell based models for myeloid leukaemia are based on mathematical models of the normal blood generation process, called haematopoiesis. 
The role of stem cells in cancer was recently reviewed in \cite{stiehl2019characterize} in terms of mathematical models which can characterise cell behaviour in normal cell development. For blood cells, an important haematopoiesis model was proposed by Marciniak et al \cite{Marciniak-Czochra2009}. The main assumption of this model was that the process of differentiation, i.e., the ability of a cell to change from one type to another, was described in several discrete maturation stages, beginning with stem cells as the first stage of maturation. 

\begin{figure}[ht]
	\centering	\includegraphics[width=0.8\textwidth]{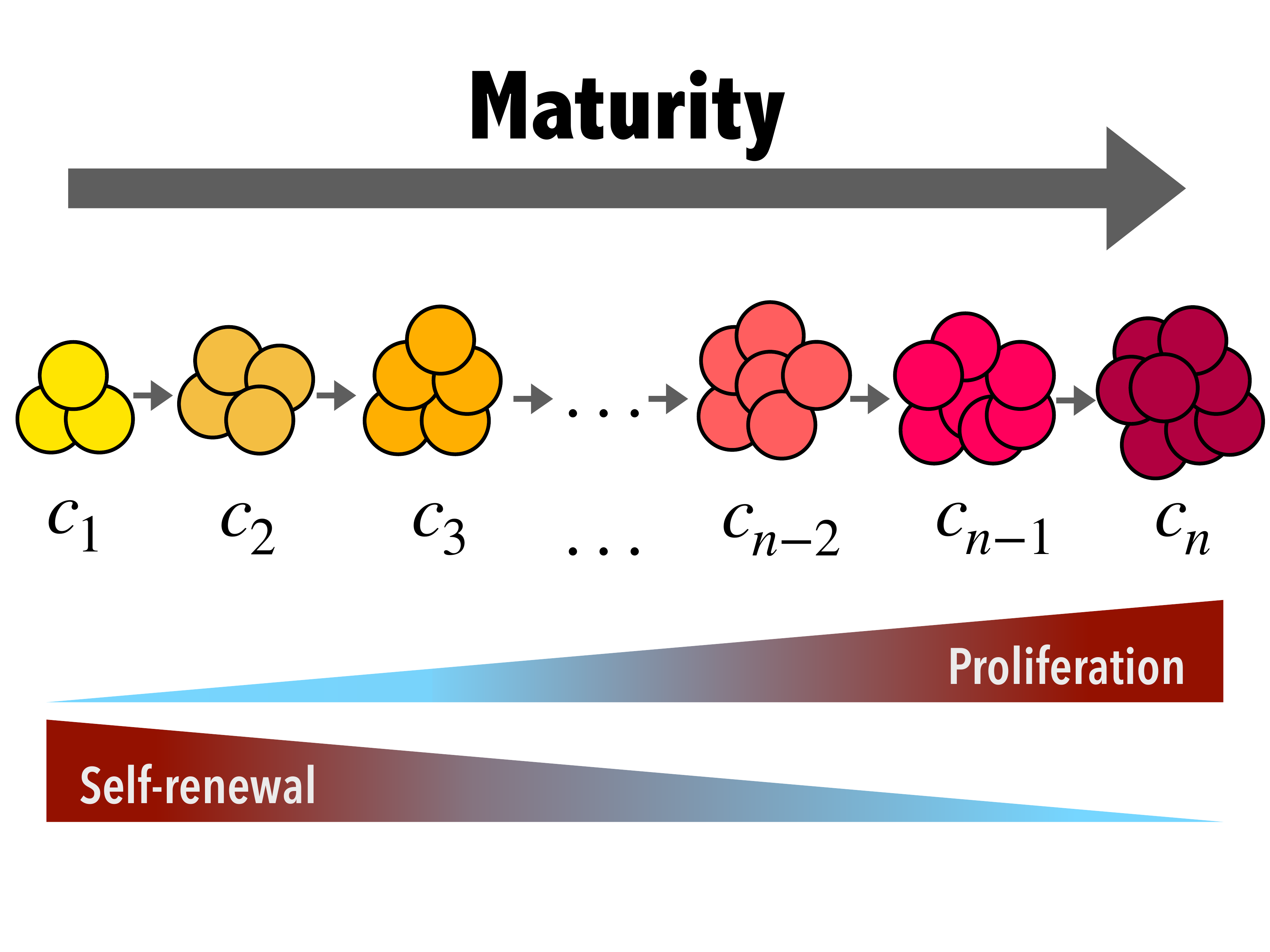}
	\caption{Schematic representation of the assumptions behind the model \eqref{MarciniakNew}. Cells were grouped into $n$ different maturation stages,  $c_i$ with $i=1,...,n$. As cells mature, and the $j$ index increases, their proliferation rates $p_j^c$ increase, whereas the self-renewal fractions $a_j^c$ decrease.}
	\label{Figure Maturity}
\end{figure}

As cells mature, their proliferation rate increases, while the self-renewal fraction lowers, where self-renewal was understood as the probability of having the same properties and fates as their parent cell. This process is summarised in Fig. \ref{Figure Maturity}. The model includes different cell subpopulations with $n$ different maturation stages and feedback signalling to regulate haematopoiesis.

The mathematical model describing the dynamics comprises a set of ODEs for the several compartments of normal cells ($c_i$) 
\begin{subequations}
	\label{MarciniakNew}
	\begin{flalign}
		\dfrac{d}{dt} c_1(t)=&(2a^c_{1,\text{max}} s(t)-1)p_1^c c_1(t)-d_1^cc_1(t),\vspace{5pt}\\
		\dfrac{d}{dt}c_i(t)=&2(1-a_{i-1,\text{max}}^c s(t))p_{i-1}^c c_{i-1}(t)+\\
		&+(2a_{i,\text{max}}^c s(t)-1)p_i^cc_i(t)-d_i^cc_i(t),\nonumber\vspace{5pt}\\
		\dfrac{d}{dt}c_n(t)=&2(1-a_{n-1,\text{max}}^c s(t))p_{n-1}^c c_{n-1}(t)-d_n^c c_n(t),
	\end{flalign}
\end{subequations}
and another set of ODEs for the leukaemic cells ($l_j$)
\addtocounter{equation}{-1}
\begin{subequations}
	\begin{flalign}
		\addtocounter{equation}{3}
		\dfrac{d}{dt} l_1(t)=&(2a^l_{1,\text{max}} s(t)-1)p_1^l l_1(t)-d_1^ll_1(t),\vspace{5pt}\\
		\dfrac{d}{dt}l_j(t)=&2(1-a_{j-1,\text{max}}^l s(t))p_{j-1}^l l_{j-1}(t)\\
		&+(2a_{j,\text{max}}^l s(t)-1)p_j^ll_j(t)-d_j^ll_j(t)\nonumber,\vspace{5pt}\\
		\dfrac{d}{dt}l_m(t)=&2(1-a_{m-1,\text{max}}^l s(t))p_{m-1}^l l_{m-1}(t)-d_m^l l_m(t),
	\end{flalign}
\end{subequations}
where $c_i=c_i(t)$ denotes the density (or number) of healthy cells in each maturation stage $i=1,..,n$,   $p_i^c$ are the proliferation rates of healthy haematopoietic cells in mitosis, $a_{i,\text{max}}^c$ are the self-renewal fractions, and $d_i^c$ the death rates for every cell maturation stage. The notation is analogous for the leukaemic cells, $l_j=l_j(t)$  for $j=1,...,m$, and the constants $p_j^l$, $a_{j,\text{max}}^l$ and $d_j^l$ for $j=1,...,m$.

\begin{figure*}[!ht]
	\centering \includegraphics[width=\textwidth]{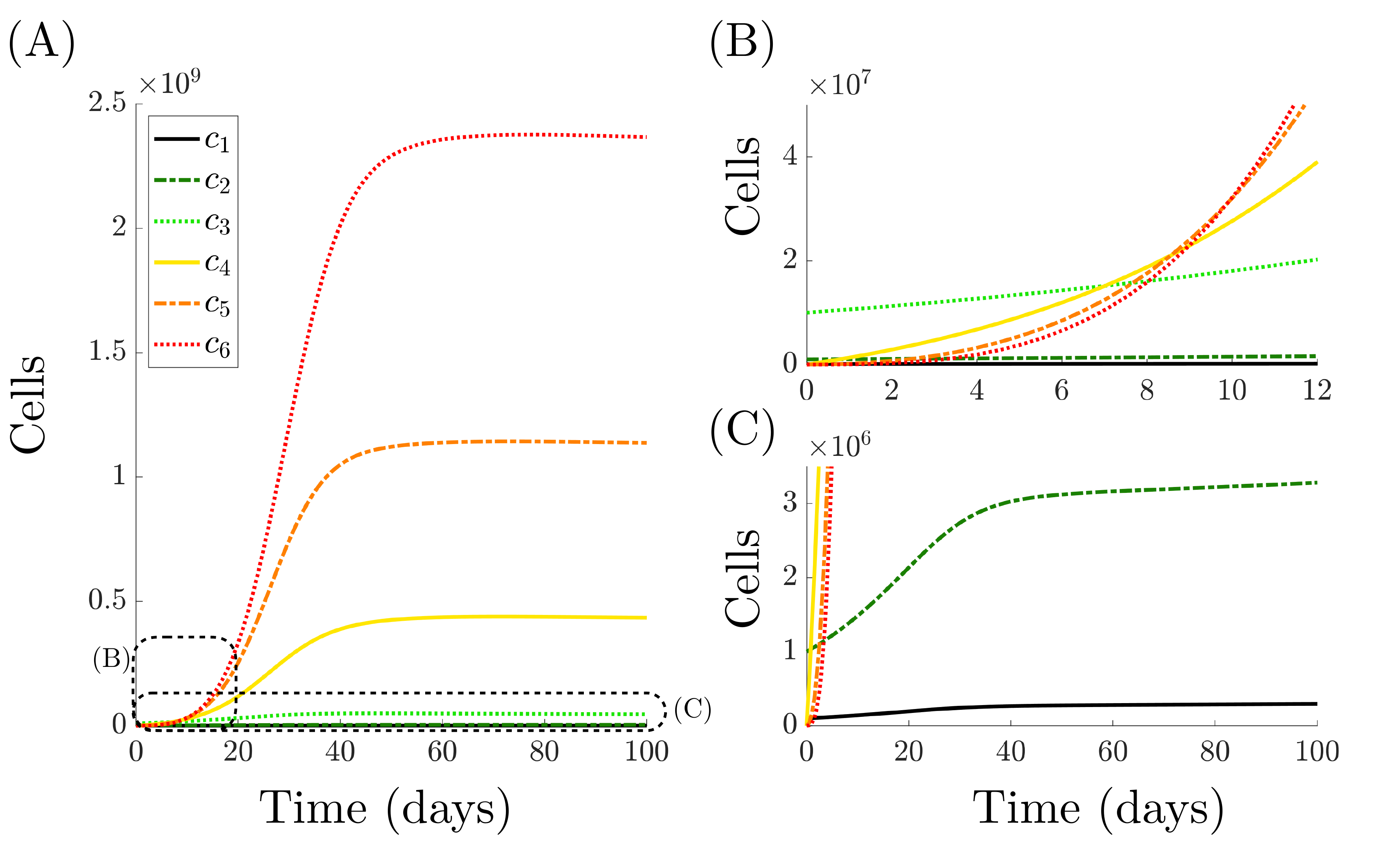}
	\caption{Simulations of the evolution of a set $n=6$ compartments accounting for six maturation stages according to the model  \eqref{MarciniakNew}. The insets A, B show more the details of the dynamics of the same simulation.  Following \cite{Marciniak-Czochra2009}, the parameter values were $a_1=0.0865$, $a_2=0.1155$, $a_3=0.1735$, $a_4=0.3465$ and $a_5=0.693$ for the self-renewal fractions. For the proliferation rates, $p_1=0.7$, $p_2=p_3=p_4=0.65$ and $p_5=0.55$ were considered. The death rate was $d=0.3$ and signal strength $k^c=1.6\cdot10^{-10}$ cells\textsuperscript{-1}. Cell initial values were $c_1(0)=10^5$, $c_2(0)=10^6$ and $c_3(0)=10^7$ and null for the other initial values.}
	\label{Fig simulations Marciniak}
\end{figure*}

Feedback signalling was described in that study using the cytokine effect function $s(t)$. Cytokines are small proteins which assist in regulating fraction chemical signalling in cells. Cytokine concentration is modelled by the equation
\addtocounter{equation}{-1}
\begin{subequations}
	\setcounter{equation}{6}
	\begin{flalign}
		\label{cytokine s}
		s(t)&=\dfrac{1}{1+k^c\,c_n(t)+k^l\,l_m(t)},
	\end{flalign}
\end{subequations}
where $k^c$ and $k^l$ are the signalling regulation strength, for both normal and leukaemic cells, respectively. These parameters are sensitive to the number of mature healthy and leukaemic cells, $c_n(t)$ and $l_n(t)$. This signalling was assumed to control the dynamics of cell proliferation and differentiation in the mathematical model. Fig. \ref{Fig simulations Marciniak} shows an example of evolution towards the homoeostatic equilibrium of the healthy haematopoietic cell compartments for $n=6$.

The model of Eqs. \eqref{MarciniakNew} in \cite{Stiehl2015} was built on the basis of the haematopoiesis model of \cite{Marciniak-Czochra2009}. The main conclusion of the mathematical study of \cite{Stiehl2015} was that both self-renewal fractions and proliferation rates could be indicators of poor prognosis. Similar models were also studied in \cite{Stiehl2012}, where some mathematical properties, including linear stability analysis, and necessary and sufficient conditions for the expansion of malignant cell clones, were studied for related models. 

A similar model by the same group \cite{Stiehl2014} described the differentiation process as a two-stage process, but considered instead the  multi-clonal nature of leukaemia, the feedback processes and the role of treatment. The study performed numerical simulations for `in-silico' virtual patients, and obtained estimated parameters from the tumour growth data of two real patients. The researchers concluded that self-renewal might be a key mechanism in the clonal selection process.  It was also stressed that late relapses could originate from clones that were already present at diagnosis, a question that has been the subject of discussion in the biomedical research literature. Stem cell self-renewal has been reviewed in terms of their impact on the dynamics of cell populations in \cite{stiehl2017stem}, concluding that a high self-renewal fraction can lead to faster cancer growth.

A similar model,  \cite{Stiehl2016}, accounted for genetic instability through the inclusion of the possibility of mutations, an essential hallmark in cancer evolutionary dynamics. Through comparison of patient data and simulations, the authors highlighted the fact that the self-renewal potential of the first emerging leukaemic clone would have a major impact on the emergence of clonal heterogeneity so that it might serve as a biomarker of patient prognosis. A recent study of the group \cite{Lorenzi2019} on acute leukaemias formalised the clonal selection dynamics via integro-differential equations. They concluded that clonal selection was driven by the self-renewal fraction of Leukaemic Stem Cells (LSCs), constructing numerical solutions based on patient data parameters from the existing literature. These simulations showed that high self-renewal for LSC clones was a marker of stability in the presence of interclonal heterogeneity. 

The model set out in Eq. \eqref{MarciniakNew} was further used in \cite{Walenda2014}  to study feedback signals from myelodysplasic syndrome (MDS) clones and their effect on normal haematopoiesis. The model was fitted using serum samples from 57 MDS patients and five healthy controls. On the basis of the numerical simulations, the authors reached the conclusion that a high self-renewal fraction of MDS-initiating cells may be critical for the development of the disease. It was conjectured that remission could be achieved if this parameter could be lowered.

Considering the dependence of leukaemic cell to cytokines, the model \eqref{MarciniakNew} is compared in \cite{stiehl2018mathematical} to a mathematical model including cytokine-independent leukaemic cell proliferation. In it, leukaemic cells are not controlled by cell signalling as in Eq. \eqref{cytokine s}, but instead a death rate is included that increases with the number of cells in the bone marrow, and acts on all cell types residing in bone marrow. This allows the authors to explain unexpected responses in some patients, such as blast crises or remission without chemotherapy. This was done by assigning patient data to two different groups that differ with respect to overall survival: those with cytokine-dependent or cytokine-independent leukaemic cell populations.

\subsection{Cell-cycle-based mathematical models of myeloid leukaemias.}
\label{Section Cell cycle leukaemia}
In some CML patients, symptoms may recur \cite{Milton1989}. This is why periodicity is specifically studied for this disease. Thus, several authors considered the cell cycle in order to explain periodicity. 

\begin{figure}[!ht]
	\centering\includegraphics[width=0.8\textwidth]{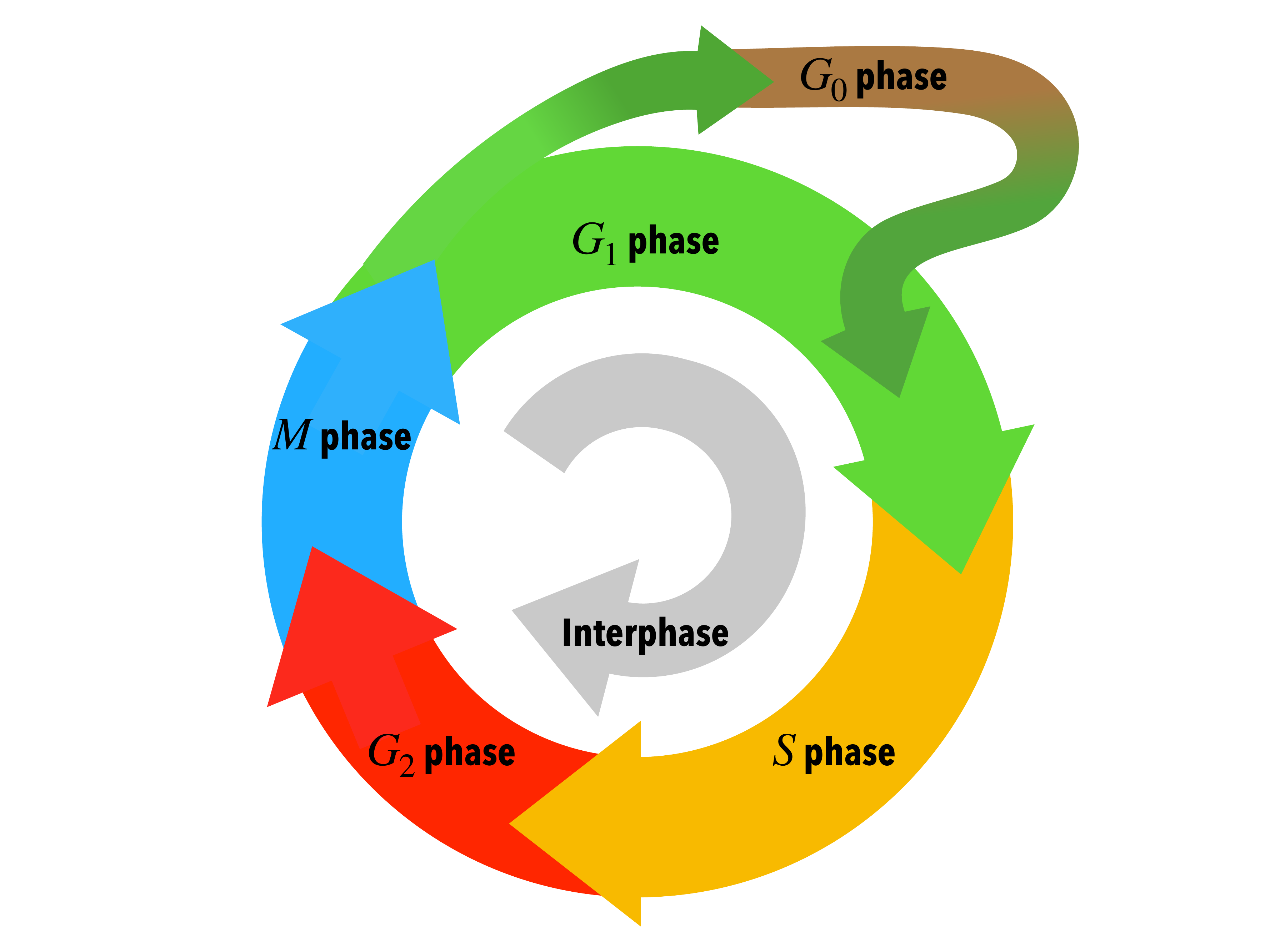}
	\vspace{5pt}
	\caption{Overview of the cell cycle. The $G_1$ phase prepares for DNA replication and synthesis in the $S$ phase. $G_2$ prepares cells for the mitosis phase $M$. While in $G_1$, cells can become quiescent, entering a resting phase $G_0$.}\label{Figure Cell Cycle}
\end{figure}

The cell cycle is the process regulating cell division. It is a multi-stage process including, firstly, mitosis ($M$), the process of nuclear division; and a stage called interphase, the interlude between two $M$ phases. In the interphase, three different substages occur: the $G_1$ phase, in which the cell prepares DNA synthesis; the $S$ phase, where DNA replicates; and the $G_2$ phase, where the cell prepares for mitosis. Any cell, before going the $S$ phase, can enter a resting state called $G_0$, where the cell becomes quiescent and remains in a non-proliferating stage. This process is summarised in Fig. \ref{Figure Cell Cycle}.

Many mathematical models have considered different aspects of the cell cycle \cite{weis2014data}. However, many of those models, arising from the so-called systems biology approach,  are quite complex. Due to the periodic nature of the cell cycle in proliferating cell populations, several mathematical models have tried to account for this cycling behaviour in a simplified form. Periodicity and other dynamic behaviours of haematological diseases are reviewed in \cite{foley2009dynamic}. 

Specifically, the model in \cite{Mackey1978} described the dynamics of blood pluripotential stem cells.  Their approach was to write equations for a population $N(t)$ of cells in the resting phase $G_0$ and another $P(t)$ of proliferating cells, described by
\begin{subequations}\label{ModelMackey}
	\begin{flalign}
		\dfrac{d N}{dt}&=-\delta N -\beta(N)N+2\beta (N_\tau)N_\tau e^{-\gamma \tau},\quad \tau<t\vspace{5pt}\\
		\dfrac{dP}{dt}&=-\gamma P +\beta(N)N-\beta (N_\tau)N_\tau e^{-\gamma \tau},\quad \tau<t,
	\end{flalign}
\end{subequations}
where $N_\tau=N(t-\tau)$, $\tau$ being the cell cycle time. The function $\beta(N)  =  \beta_0/\left(1+(N/N_*)^n\right)$ is the mitotic re-entry rate, i.e. the rate of cell entry into proliferation, where $\beta_0$, $ N_*$, $n$ are parameters. The parameter $\delta$ is the total differentiation fraction from the $G_0$ phase, and $\gamma$ is the fraction of irreversible cell loss from all portions of the proliferating-phase stem-cell population. Taking values for these parameters from the literature, the authors concluded that the origin of aplastic anaemia and periodic haematopoiesis could be related to irreversible cell loss from the blood pluripotential stem compartment.  

\cite{Mackey2006} studied Eqs. \eqref{ModelMackey}, to describe the existence and stability of long-period oscillations of stem cell populations in periodic chronic myelogenous leukaemia.  This was made possible by studying a contractive return map, such that a fixed point of the return map gave a stable periodic solution of the model equation. This was computed in such a way that there was no analytic formula for the periodic solution in the limiting case $n\rightarrow\infty$.

Other work based on the \eqref{ModelMackey} model, such as \cite{Colijn2005a}, gives estimates of the model parameters for a typical normal human, and explored the changes in some of these parameters necessary to account for the quantitative data on leukocyte, platelet and reticulocyte cycling in 11 patients with Periodic Chronic Myelogenous leukaemia (PCML). Their results indicated that the critical model parameter changes required to simulate the PCML patient data were an increase in the amplification in the leukocyte line, an increase in the differentiation fraction from the stem cell compartment into the leukocyte line, and the rate of apoptosis in the stem cell compartment. In a companion study \cite{Colijn2005}, they found that the parameter changes that mimic untreated cyclical neutropenia correspond to a decreased amplification (increased apoptosis) within the proliferating neutrophil precursor compartment, and a decrease in the maximal rate of re-entry into the proliferative phase of the stem cell compartment. The case of granulocyte colony stimulating factor treatment was also studied. Safarishahrbijari and Gaffari \cite{Safarishahrbijari2013} used the equations for red blood cells and platelets from \cite{Colijn2005a} and for leukocytes from \cite{foley2009dynamic} to identify parameters in PCML. The inclusion of new parameters resulted in a better fit of clinical data and from the data extracted from both platelet and leukocyte models.

Pujo-Menjouet and Mackey \cite{Pujo-Menjouet2004}, performed a local stability analysis of the model \eqref{ModelMackey} and found the conditions for Hopf bifurcation to occur. Periodic oscillations were studied depending on five haematopoietic stem cell parameters: the mitotic rate sensitivity, the maximal rate of cell entry into proliferation from the resting $G_0$ phase, the differentiation and apoptosis rate and the time to entry into mitosis. Extensions of this work \cite{Halanay2012}, have proven that, under periodic treatment, there is a periodic solution with the same period. This could be related to the observed oscillatory behaviour of blood cells' counts under treatment in CML. 

\begin{figure}[ht]
	\centering	\includegraphics[width=0.8\textwidth]{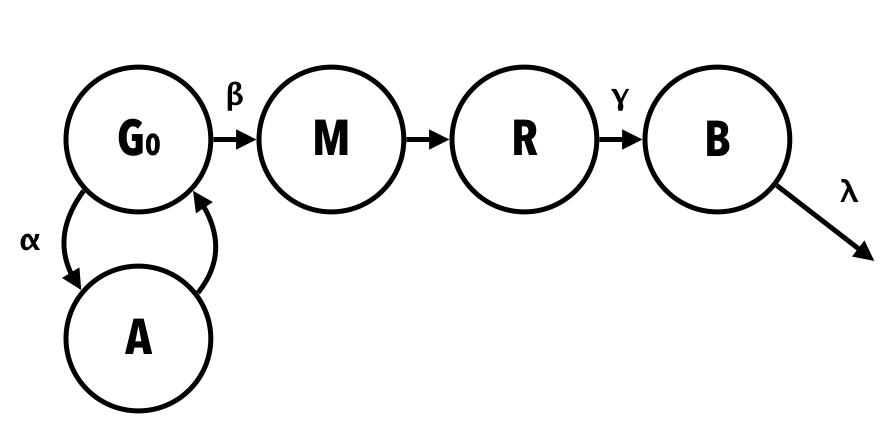}
	\caption{Schematic view of the cell compartments in \cite{Rubinow1976}. $A,G_0,M,R$ and $B$ represent different cell compartments, $\alpha,\beta,\gamma$ are the corresponding coupling rates between them, and $\lambda$ is the irreversible blood cell loss.}\label{Rubinow}
\end{figure}

A different type of models to describe myeloblastic leukaemias have been constructed on the basis of the work of Rubinow and Lebowitz \cite{Rubinow1976}. The model itself was based on granulocytopoiesis, also studied by these authors in \cite{Rubinow1975}. In this first work, qualitative analysis was performed, supporting evidence for alterations which presumably occurs in cyclic neutropenia. For both models they considered four compartments for healthy cells as shown schematically in Fig. \ref{Rubinow}: the active $A$ and $G_0$ cell compartments, representing the proliferative pools, and the maturation $M$  and reserve $R$ cell compartments, which finally ended in the blood pool $B$. For the leukaemic cells, only active and $G_0$ cells were considered, of which only a certain fraction were released into the blood, with no further maturation stages.

For this model, and in terms of myeloid leukaemia, the presence of a leukaemic population destabilises the homoeostatic state of the normal population, which is stable in the absence of leukaemic cells. In \cite{Rubinow1976a}, the authors found differences between normal and leukaemic cell populations but including treatment into the model from Fig. \ref{Rubinow}: firstly, the recovery rate was higher for normal cells, as compared to leukaemic cells from the action of cytotoxic treatment. Secondly, the $S$-phase duration was different for the two populations. This led the authors to the conclusion that, for patients with a ``slow'' growing leukaemic cell population, remission could be achieved with one or two courses of treatment, whereas for those with a ``fast'' growing leukaemic cell population, a similar aggressive treatment achieved remission only at the cost of great toxic effects on the normal cell population.

These mathematical models described both the processes of normal blood and myelogenous leukaemia development. \cite{Fokas1991} did the same, but for CML.  The authors showed how CML cells could ultimately outnumber normal cells. They used the model to study the relationship between proliferation and maturation and proposed a solution to the apparent contradiction between decreased proliferation and increased production, by assuming that a greater fraction of CML cells is produced by division rather than by maturation.

Another mathematical model \cite{Fuentes-Gari2015} included details on cyclins D, E and B,  a family of proteins that help to control the cell cycle. Their production has a direct influence on the transition of a cell in the $G_0$, $G_1$ and $G_2$ phases, respectively. Flow-cytometry data profiles for three leukaemia cell lines were analysed in this study (K-562, MEC-1, and MOLT-4, from AML, CLL and ALL patients, respectively).  For the $S$ phase, DNA replication was considered, as it is key before a cell can produce new daughter cells.  The authors assumed that $G=G(C_E,t)$ was the number of cells in $G_0/G_1$ at time $t$ with a cyclin E content $C_E$. Similarly, they denoted by $S=S(DNA,t)$ and $M=M(C_B,t)$ the number of cells in the $S$ and $G_2/M$ phases that had DNA content (represented as the variable $DNA$) and cyclin B content $C_B$ at time t, respectively. These assumptions taken together led to the model
\begin{subequations}
	\label{Model Fuentes}
	\begin{flalign}
		&\dfrac{\partial G}{\partial t}+\dfrac{\partial\left(G\cdot \dfrac{d C_E}{dt}\right)}{\partial C_E}=-r_{G\rightarrow S}(C_E)\cdot G,\vspace{5pt}\\
		&\dfrac{\partial S}{\partial t}+\dfrac{\partial\left(S\cdot \dfrac{d DNA}{dt}\right)}{\partial DNA}=0,\vspace{5pt}\\
		&\dfrac{\partial M}{\partial   t}+\dfrac{\partial\left(M\cdot \dfrac{d C_B}{dt}\right)}{\partial C_B}=-r_{M\rightarrow G}(C_B)\cdot M,\vspace{5pt}
	\end{flalign}
\end{subequations}
where $r_{G\rightarrow S}$, $r_{M\rightarrow G}$ are the transition fractions from $G_2/M$ to $G_0/G_1$ and from $G_0/G_1$ to the $S$ phase, respectively. Good agreement was found between experimental results and the model simulations.  This could assist in developing clonal models of leukaemogenesis. The authors claimed that the model could help in the identification of heterogeneous leukaemia clones at diagnosis and post-treatment, and that it could have the potential to predict future outcomes in response to induction and consolidation chemotherapy as well as relapse kinetics.

\subsection{Other data-based mathematical models of myeloid leukaemia.}
\label{Section Acute and Chronic leukaemia}

Myeloid leukaemia models are the most studied in the literature. \cite{Sarker2017}, for example, describes acute myeloid leukaemia (AML) using a multi-lineage multi-compartment model of the haematopoietic system and feedback via cytokines and chemokines. Analysis of the model suggested that self-renewal probabilities,  mitotic rates and cytokine growth factors produced in peripheral blood determined leukocyte homoeostasis. The mitosis rate of cancer was found to be the parameter with the strongest prognostic value. 

A comparison of three mathematical models that describe CML progression and aetiology was undertaken in \cite{MacLean2014}. The authors sought to identify which models could provide the best description of disease dynamics and their underlying mechanisms. The first considered the following dynamic system
\begin{subequations}\label{Model Maclean}
	\begin{flalign}{}
		\dfrac{dx_0}{dt}&=a_x x_0(k-z)-b_x x_0,\vspace{5pt}\\
		\dfrac{dx_1}{dt}&=b_x x_0 +c_x c_1(k-z)- d_x x_1,\vspace{5pt}\\
		\dfrac{dx_2}{dt}&=d_x x_1 - e_x x_2,\vspace{5pt}\\
		\dfrac{dy_0}{dt}&=a_y y_0(k-z)-b_y y_0,\vspace{5pt}\\
		\dfrac{dy_1}{dt}&=b_y y_0 - e_y y_1,
	\end{flalign}
\end{subequations}
where $x_0$ were HSCs, $x_1$ healthy progenitors, $x_2$ differentiated cells, $y_0$ LSCs and $y_1$ differentiated leukaemic cells, with parameters $a$, $b$, $c$, $d$, $e$ as the corresponding self-renewal, production and death rates. Also, $k$ was the carrying capacity and $z=x_0+x_1+x_2+y_1+y_2$ the total number of cells.  The second model, in \cite{Michor2005}, was a shorter version of the \eqref{Model Michor} model, to be presented in detail later. The third was \cite{Foo2009}, which allowed competition between HSC and LSCs. This latter model was based on the following ODEs:
\begin{equation}\label{Model Foo}
\begin{array}{lllllllll}
\begin{array}{lllllllll}
\dfrac{dx_0}{dt}&=\beta_x x_q+&\dfrac{dy_0}{dt}&=\beta_y y_q+\vspace{5pt}\\
&\hspace{-5pt}+(r_x\phi_x-d_0-\alpha_x)x_0,&&\hspace{-5pt}+(r_y\phi_y-d_0-\alpha_y)y_0,\\
\dfrac{dx_q}{dt}&=\alpha_x x_0 - \beta_x x_q&\dfrac{dy_q}{dt}&=\alpha_y y_0 - \beta_y y_q\vspace{5pt}\\
\dfrac{dx_1}{dt}&=a_x x_0 - d_1 x_1,&\dfrac{dy_1}{dt}&=a_y y_0 - d_1 y_1,\vspace{5pt}\\
\dfrac{dx_2}{dt}&=b_x x_1 - d_2 x_2,&\dfrac{dy_2}{dt}&=b_y y_1 - d_2 y_2,\vspace{5pt}\\
\dfrac{dx_3}{dt}&=c_x x_2 - d_3 x_3,&\dfrac{dy_3}{dt}&=c_y y_2 - d_3 y_3.\\
\end{array}
\end{array}
\end{equation}
The healthy cells $x_i$ and leukaemic cells $y_i$ were considered at different stages $i=0,...,3$ of differentiation and a compartment of quiescent cells was also added for each type, $x_q$ and $y_q$. The authors found that it was not possible to choose between the models based on fits to the data of 69 patients who had experienced relapse or remission of the disease. They suggested experiments directly probing the haematopoietic stem-cell niche that could help in choosing the best model.

Finally, \cite{Michor2006} described another model of cancer initiation for CML. The authors assumed that the clonal expansion of mutant cells is given by a logistic equation
\begin{equation}
\dfrac{dx}{da}=\dfrac{r-1}{\tau}x(1-x), \qquad \text{with } x(0)=\dfrac{1}{N},
\end{equation}
where $a$ is the time since mutation happened, $x(a)$ the frequency of mutant clones with $r$ relative fitness, and $N$ the total cell population with generation time $\tau$. $q$ was the rate of detection and $u$ the probability per cell division of producing a mutant cell. Letting $c=\frac{r-1}{\tau}$, and $b=N\, u\,\frac{c}{r}$, the probability of detecting cancer before time $t$ was given by
\begin{equation}
P(t)=\int_0^t b\,\exp(-b\,m)\left(1-\exp(H(t,m))\right)\, dm,
\end{equation}
for 
\begin{equation}
H(t,m)=-q\,N\int_{0}^{t-m}\dfrac{da}{1+(N-1)\exp(-c\,a)},
\end{equation}
with $m$ a small probability that the first mutant arose early. Interestingly, this simple model, based only on the Philadelphia translocation, gave rise to cancer incidence curves with exponents of up to 3 as a function of age. This behaviour had been previously thought to be associated necessarily with three mutations, two of which were unknown. Thus, the model proved that CML incidence data were consistent with the hypothesis that the Philadelphia translocation alone could cause CML.

\subsection{Other theoretical studies of myeloid leukaemias}
\label{Section General Models leukaemia}

\begin{figure}[ht]
	\centering 	\includegraphics[width=0.8\textwidth]{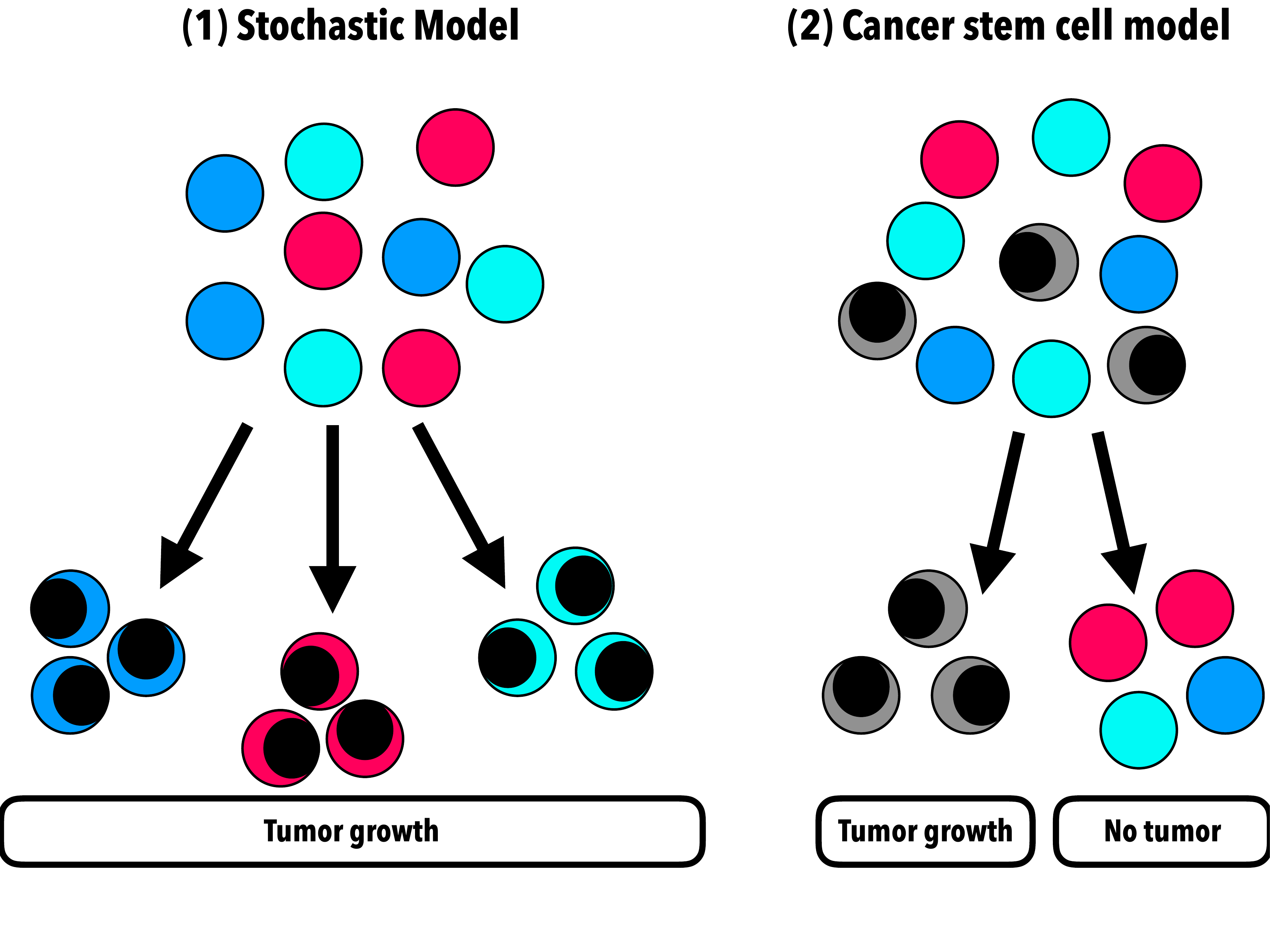} 
	\caption{Different representations of tumour proliferation models. In the stochastic model, tumour cells are heterogeneous, so that genetic changes that lead to tumour cells can originate from any cell. In the CSC model, only a small subset of cancer cells has the ability to initiate new tumour growth. Fig. adapted from Ref. \cite{Wang2005}.} 
	\label{Figure Model Stem Cells}
\end{figure}

Cancer initiation and maintenance are typically assumed to be related to cancer stem cells (CSCs)
\cite{Chaffer2015,Visvader2011}. Two models for cancer initiation have been derived using this assumption. The first is a genetic mutation model, where mutations determine the phenotype of the tumour. In this conceptual framework different mutations may result in different tumour morphologies, even when starting from the very same stem cell. Cells inherit the molecular alteration and regain the ability for self-renewal, which leads to a population of cancer cells. The second model assumes that different cells serve as cells of origin for the different cancer subtypes, the so-called CSCs. This model proposes that oncogenic events occur in different cells, and these produce different kinds of cancer.  In this model, self-renewal potential is limited for the CSCs. Both conceptual models are shown schematically in Fig. \ref{Figure Model Stem Cells}.

\cite{Komarova2005} constructed a stochastic model that considered drug resistance for CML, where the probability of treatment failure was approximated by
\begin{equation}
M _ { 0 } \frac { n ! ( L - D ) L ^ { n - 1 } u ^ { n } } { ( D + H - L ) ^ { n } },
\end{equation}
for $M_0$ the initial non-mutant cells, $n$ the quantity of drugs used, $u$ the probability of mutation after cell division, and finally, the measurable parameters $L$, $D$ and $H$ as, respectively, the rate of growth, death and the drug-induced death rate. From the analysis of the mathematical model, the authors claimed that although drug resistance prevented successful treatment, resistance could be overcome with a combination of three targeted drugs.

Finally, several authors have built models of leukaemias using graph-theoretical methods. 
Graphs can be used to describe the hierarchical organisation observed in haematopoiesis, as seen in Fig. \ref{Figure haematopoiesis}. \cite{Cho2018} parametrised a graph-theoretical model of haematopoiesis using publicly available RNA-Seq data in a high-dimensional space. The high-dimensional data were later reduced to $\mathbb{R}^2$ or $\mathbb{R}^3$ using reduction techniques, such as principal component analysis, diffusion maps and t-distributed stochastic neighbour embedding, and a PDE model on a graph $G$ was constructed. $u(x,t)$ denoted the cell distribution at the differentiation continuum space location $x\in G$ and time $t$. Then, for every cell distribution $u_k(x,t)$ on an edge $e_k$, the cell density was modelled with advection-diffusion-reaction equations
\begin{equation}
\frac { \partial u _ { k } } { \partial t } = R _ { k }  u _ { k }  - \frac { \partial \left( V _ { k } u _ { k } \right)} { \partial x }  + \frac { D _ { k } } { 2 w _ { k }  } \frac { \partial } { \partial x } \left( w _ { k } \frac { \partial u _ { k } } { \partial x } \right) ,
\end{equation}
for $ x \in e _ { k } = \overline { a _ { k } b _ { k } },$ where each edge $e_k$ was parametrised from $a_k$ to $b_k$, and the following functions were considered:  $R_k=R_k(x)$ as the cell proliferation, $V_k=V_k(x)$ as the advection coefficient, and apoptosis and diffusion terms $D_k=D_k(x)$ and $w_k=w_k(x)$, which respectively describe cell fluctuation and width of a narrow domain around an edge. Using this model, the authors performed simulations consistent with the evolution of AML populations. A similar approach was used in \cite{Daniel2002}, where the graphs constructed presented the essential properties of functioning bone marrow.

\section{Mathematical description of Chronic Myeloid Leukaemia Treatments.}
\label{Section treatment CML} 


\subsection{Imatinib and its basic mathematical modelling}

CML has been intensively studied in terms of therapy based on Imatinib. This drug is a 2-phenyl amino pyrimidine derivative that inhibits a number of tyrosine kinase (TK) enzymes. Imatinib is specific for the TK domain in ABL (the Abelson proto-oncogene), c-kit and PDGF-R (platelet-derived growth factor receptor).
In chronic myelogenous leukaemia, the Philadelphia chromosome leads to a fusion protein of ABL with the breakpoint cluster region, termed BCR-ABL. Imatinib decreases the BCR-ABL activity. CML treatments have been strongly influenced by the appearance of imatinib \cite{Deininger2005}, that is now the standard first-line treatment against the disease. It is a very effective drug with up to about 70\% of people having a complete cytogenetic response (CCyR) within 1 year of starting imatinib. After a year, even more patients will have had a CCyR. Many of these patients also have a complete molecular response (CMR).

The capacity of the drug to impair the proliferation of leukaemic stem cells was the basic assumption behind the mathematical model of Michor and co-workers \cite{Michor2005}. The model also included  the development of resistance to therapy and 
was based on the following system of differential equations:
\begin{equation}\label{Model Michor}
\arraycolsep=2.7pt
\begin{array}{c}
\begin{array}{lllllll}
\dfrac{dx_0}{dt}=(\lambda(x_0)-d_0)x_0,&\dfrac{dy_0}{dt}=(r_y(1-u)-d_0)y_0,\vspace{5pt}\\
\dfrac{dx_1}{dt}=a_x x_0 - d_1 x_1,&\dfrac{dy_1}{dt}=a_y y_0 - d_1 y_1,\vspace{5pt}\\
\dfrac{dx_2}{dt}=b_x x_1 - d_2 x_2,&\dfrac{dy_2}{dt}=b_y y_1 - d_2 y_2,\vspace{5pt}\\
\dfrac{dx_3}{dt}=c_x x_2 - d_3 x_3,&\dfrac{dy_3}{dt}=c_y y_2 - d_3 y_3,\vspace{5pt}\\
\end{array}\\
\begin{array}{lllllll}
\dfrac{dz_0}{dt}=(r_z-d_0)z_0+r_yy_0u,\vspace{5pt}\\
\dfrac{dz_1}{dt}=a_z z_0 - d_1 z_1,\vspace{5pt}\\
\dfrac{dz_2}{dt}=b_z z_1 - d_2 z_2,\vspace{5pt}\\
\dfrac{dz_3}{dt}=c_z z_2 - d_3 z_3.\vspace{5pt}\\
\end{array}
\end{array}
\end{equation}
Here, $x_i$ denotes the different populations of normal cells, $y_i$ the imatinib-sensitive leukaemic populations and $z_i$  the tumour clones resistant to imatinib. The indexes $i=0,1,2,3,$ denote the subpopulations of stem cells, progenitors, differentiated and terminally differentiated cells in each compartment. The rate constants for each cell type ($x,y,z$) are given by $a,b$ and $c$,  and $d_i$ are the death rates for $i=0,1,2,3$. 
Cell division rates are given by $r_y$ and $r_z$. The parameter $u$ is the fraction of resistant cells produced per cell division. Finally, $\lambda=\lambda(x_0)$ is a decreasing function of $x_0$ describing homoeostasis of normal stem cells. It models the feedback signals controlling haematopoiesis. Data from 169 CML patients were used to fit the mathematical model in \cite{Michor2005}. The authors obtained numerical estimates for the turnover rates of leukaemic progenitors and differentiated cells and showed that imatinib dramatically reduced the rate at which these cells are being produced from leukaemic stem cells. They showed that the probability of harbouring resistance mutations increases with disease progression as a consequence of an increased leukaemic stem cell abundance, and proposed that the time to treatment failure caused by acquired resistance is given by the growth rate of the leukaemic stem cells. Their bottom line was that multiple drug therapy is especially important for patients who are diagnosed with advanced and rapidly growing disease.


A simplified version of the model \eqref{Model Michor} was studied in \cite{Dingli2006} by considering only the stem cell (0) and differentiated cell (1) compartments of healthy ($x$) and leukaemic ($y$) cells, i.e.
\begin{subequations}\label{Model Dingli}
	\begin{eqnarray}
	\dfrac{dx_0}{dt} & = & (r_x \phi - d_0) x_0, \\
	\dfrac{x_1}{dt} & = & a_x x_0 - d_1 x_1, \\
	\dfrac{y_0}{dt}  & = & (r_y \varphi - d_0) y_0, \\
	\dfrac{y_1}{dt} & = & a_y y_0 - d_1 y_1, 
	\end{eqnarray}
\end{subequations}
where $\phi=1 /\left[1+c_{x}\left(x_{0}+y_0\right)\right]$ and $\psi=1 /\left[1+c_{y}\left(x_{0}+y_{0}\right)\right]$ are homeostasis functions for normal and tumour stem cells respectively, and $c_x,c_y$ are Michaelis-Menten parameters. By a combination of analysis and simulation, the authors discussed how any successful therapy would require the eradication of the pool of leukaemic stem cells; otherwise, progressive disease is very likely. Thus, successful therapeutic agents must enhance the death rate of this rare population of cells. Therapies designed to target mitosis of malignant stem cells could not eradicate the disease quickly. Nevertheless, there has been some controversy surrounding the potential effectiveness of imatinib to achieve remission \cite{Michor2007}.

In \cite{Kim2008}, the immune response targeting leukaemic cells was added to Eqs. \eqref{Model Michor}. Using experimental data  from the literature, a mathematical model was fitted in which immune response was described by delay differential equations. The authors considered that T cells targetting leukaemic cells could prevent relapse, and combine with imatinib therapy. The more simplified model in Eq. \eqref{Model Dingli} was later used by \cite{Paquin2010} to study and numerically simulate treatment interruptions as a potential therapeutic strategy for CML patients. In many cases, strategic treatment interruptions led to the elimination of leukaemic cells in silico. The conclusion was that strategic treatment interruptions could be a feasible clinical approach to enhancing the effects of imatinib treatment for CML.

A number of extensions of the \eqref{Model Michor} model have been developed for CML. For example, in \cite{Olshen2014}, four levels of cell differentiation were included and studied for the BCR-ABL1 gene, necessary for the pathogenesis of CML. In that study, data from 290 patients were used, 92 of them treated with dasatinib, 75 with nilotinib and 123 with imatinib.  All treatments elicited similar responses. Another extension of the model was described in \cite{Helal2015}, with a focus on more theoretical aspects, including a stability analysis, and an existence proof for positive solutions. 

The global dynamics of normal and CML haematopoietic stem cells and differentiated cells were also studied in \cite{ainseba2010optimal}. The dynamic was assumed to be governed by the following system of Lotka-Volterra equations
\begin{subequations}\label{Theo1}
	\begin{flalign}
		\dfrac{dx_0}{dt}&=n\left(1-\dfrac{x_0+y_0}{K}\right)x_0-d_0 x_0,\vspace{5pt}\\
		\dfrac{dx_1}{dt}&=r x_0-(d-d_2)x_1,\vspace{5pt}\\
		\dfrac{dy_0}{dt}&=m\left(1-\dfrac{x_0+\alpha y_0}{K}\right)y_0-g_0 y_0,\vspace{5pt}\\
		\dfrac{dy_1}{dt}&=q y_0-(g-g_2)y_1,
	\end{flalign}
\end{subequations}
where  $x_0(t)$ represents haematopoietic normal stem cells (HSC), $x_1(t)$ normal differentiated cells and  $y_0(t)$ and $y_1(t)$ describe the same subpopulations of cancer cells. In Eqs. (\ref{Theo1}) $n$, $m$, $r$, $q$ are division rates, $d_0$, $d$, $g_0,g$  death rates, $K$ the carrying capacity and $\alpha\in]0,1[$ is a constant. The production rates for differentiated cells are given by $d_2$ and $g_2$.
Several optimal control problems were solved for imatinib, whose effect on the division and mortality rates of cancer cells produces a suboptimal response. The effect of cyclic combination of two drugs in CML was studied in \cite{Komarova2011}, and the modelling led to the conclusion that treatments should start with the stronger drug, and the weaker one should have cycles of longer duration. 

An interaction model between naïve T cells (mature T cells from thymus), effector T cells (cells which actively respond to stimuli) and CML cancer cells was described in \cite{Moore2004}, where Latin hypercube sampling was used to estimate parameter values due to the lack of data. This is a statistical technique for generating parameters from a multidimensional distribution. In their conclusion, the authors explained that the growth rate of CML and the natural death rate were the most important parameters, suggesting that treatment for CML patients should focus on these rates. Any drug with a high cost that is included in the model could be studied in order to obtain optimal treatment, and reduce not only radiation but also financial benefits. This model was later used in \cite{Berezansky2012}, focusing on cancer $x=x(t)$ and effector $y=y(t)$ cell population dynamics, by considering a combined treatment with imatinib and the interferon-alpha (IFN-$\alpha$) therapy. This last is a protein whose activation produces a cytogenetic response in CML patients. The model considered the following ODEs
\begin{subequations}
	\begin{flalign} 
		\dfrac{dx}{dt} =&\beta_{1} x(t) \ln \frac{K}{x(t)}-\gamma_{1} x(t) y(t)-\omega \gamma_{3} x(h(t)), \\ 
		\dfrac{dy}{dt} =&\beta_{2} \frac{x(t)}{\eta_{1}+x(t)} y(t)-\gamma_{2} x(t) y(t)+\\
		&+in_{\alpha} \gamma_{4} \frac{y(t)}{\eta_{2}+y(t)} y(t-\tau)-\mu_{y} y(t), \nonumber
	\end{flalign}
\end{subequations}
where $\beta_1$, $\beta_2$ were the respective reproduction rates, $K$ the maximal tumour population, $\eta_1$, $\eta_{2}$ Michaelis-Menten terms and $\gamma_{1}$, $\gamma_{2}$ the cell loss rates due to interaction. The death rate for effector cells is $\mu_y$, while tumour death is modelled by the constants $\omega \gamma_3$ and a function $h(t)$. This function is modelled as $h(t)=t-\theta e^{-\lambda t}$, so that the influence of drugs tends to zero over time. The dose of IFN-$\alpha$ is modelled as $in_{\alpha} \gamma_{4}$, which increases the effector cell population with a delay $\tau$ of about 7 days.  The stability analysis proposed, as well as the results obtained, were able to describe the influence of two types of the treatment. The authors claimed that the dose of IFN-$\alpha$ has an inhibitory effect on the number of cancer cells, but its replacement with another type of treatment should be considered in order to avoid resistance. 

Finally, \cite{Nanda2007} studied optimal control problems for CML, in a model with a molecular targeted therapy such as imatinib. Naïve T cells, which are already differentiated T cells, but are precursors for more mature cells called effector cells, were also included in the model. The cancer cell population was then activated by the presence of the CML antigen. Aiming to minimise the cancer cell population and the detrimental effects of the drug, they found that a high dose level from the beginning was optimal. Also, combination therapy was better than single dosing.

\subsection{Modelling the effect of quiescence on Imatinib treatments}

Quiescence, which corresponds to the $G_0$ phase of the cell cycle, and its relationship to drug therapy (in this case, imatinib)  is an important factor in leukaemia because quiescent cells might not be affected by therapy, as drugs target proliferative cells, and a possible relapse may occur.

Imatinib treatment was studied using Roeder model \cite{Roeder2002,Roeder2006,Roeder2008} accounting for quiescent and proliferative cell compartments. Firstly, in \cite{Roeder2002} a stochastic model of haematopoiesis was developed. On the basis of that model, another was built to describe imatinib-treated patients \cite{Roeder2006}.

A more advanced model based on partial differential equations (PDEs) was studied in \cite{Roeder2008}.  This model considered quiescent and cycling stem cells, denoted by $A$ and $\Omega$, respectively. The authors included a cell-intrinsic function $a(t)$, which determined the affinity of a cell for residing in $A$ or  $\Omega$. With $a(t)\in [a_{\min},a_{\max}]$, a quiescent stem cell would enter the cell cycle with probability $\omega$ and a cycling cell would become quiescent with probability $\alpha$. These terms were modelled as
\begin{subequations}
	\label{Model Roeder}
	\begin{flalign}
		\omega(\Omega(t),a(t))=\frac{a_{\min}}{a(t)}f_\omega(\Omega(t)),\\
		\alpha(A(t),a(t))=\frac{a(t)}{a_{\max}}f_\alpha(A(t)),
	\end{flalign}
\end{subequations}
where the sigmoidal functions $f_\omega$ and $f_\alpha$ were defined by
\addtocounter{equation}{-1}
\begin{subequations}
	\begin{flalign}
		\setcounter{equation}{2}
		f_\omega(\Omega(t))=\frac{1}{\nu_1+\nu_2\exp\left(\nu_3\frac{\Omega(t)}{N_\omega}\right)}+\nu_4,\\
		f_\alpha(A(t))=\frac{1}{\mu_1+\mu_2\exp\left(\mu_3\frac{A(t)}{N_\alpha}\right)}+\mu_4,
	\end{flalign}
\end{subequations}
for specific values of the parameters $\nu_j$, $\mu_j$, for $j=1,2$ and the scaling factors $N_\omega$ and $N_\alpha$.  The  dynamics of the HSCs, quiescent ($A$) and proliferating ($\Omega$), were governed by these equations:
\addtocounter{equation}{-1}
\begin{subequations}
	\begin{flalign}\label{Roeder eqs 1}
		\setcounter{equation}{4}
		\dfrac{\partial n_{A}}{\partial t}+v_A\cdot \dfrac{\partial}{\partial a}n_{A}&=-\left(\dfrac{dv_{A}}{d a}-\omega\right)\cdot n_A+\alpha\cdot n_\Omega,\vspace{5pt}\\
		\label{Roeder eqs 2}
		\dfrac{\partial n_{\Omega}}{\partial t}+v_\Omega\cdot \dfrac{\partial n_{\Omega}}{\partial a}&=\left(-\dfrac{dv_{\Omega}}{d a}+\tau-\alpha\right)\cdot n_\Omega+\omega\cdot n_A.
	\end{flalign}
\end{subequations}
The functions $n_{A}=n_{A}(a,t)$ and $n_{\Omega}=n_{\Omega}(a,t)$ represent the cell densities at affinity $a$ and time $t$ within $A$ and $\Omega$, respectively. Also, $v_A=v_A(a)$ and $v_\Omega=v_\Omega(a)$ were the corresponding velocities that make $v_A\cdot n_A$ and $v_\Omega\cdot n_\Omega$ the corresponding cell fluxes for each compartment. Finally, $\tau$ was a parameter which simulates average cell division depending on cell cycle duration. 
Eqs. \eqref{Roeder eqs 1} and \eqref{Roeder eqs 2} were the basis for studying leukaemia and how the imatinib treatment affect its dynamics, in a highly efficient way when it comes to huge cell populations. They considered the dynamics for every cell subpopulation in the following system:
\addtocounter{equation}{-1}
\begin{subequations}
	\begin{flalign}\label{Roeder eqs 3}
		\setcounter{equation}{6}
		\dfrac{\partial n_{A}^{(i)}}{\partial t}+&v_A^{(i)} \dfrac{\partial n_{A}^{(i)}}{\partial a}=-\left(\dfrac{dv_{A}^{(i)}}{d a}-\omega^{(i)}\right) n_A^{(i)}+\alpha^{(i)} n_\Omega^{(i)},\vspace{5pt}\\
		\label{Roeder eqs 4}
		\dfrac{\partial n_{\Omega}^{(1)}}{\partial t}+&v_\Omega^{(1)} \dfrac{\partial n_{\Omega}^{(1)}}{\partial a}=\omega^{(1)} n_A^{(1)}+\vspace{5pt}
		\\
		&+\left(-\dfrac{dv_{\Omega}^{(1)}}{d a}+\tau^{(1)}-\alpha^{(1)}\right) n_\Omega^{(1)},\nonumber\\
		\label{Roeder eqs 5}
		\dfrac{\partial n_{\Omega}^{(2)}}{\partial t}+&v_\Omega^{(2)} \dfrac{\partial n_{\Omega}^{(2)}}{\partial a}=\omega^{(2)} n_A^{(2)}+\\
		&+\left(-\dfrac{dv_{\Omega}^{(2)}}{d a}+\tau^{(2)}-\alpha^{(2)}-r_{\text{inh}}-r_{\text{deg}}\right) n_\Omega^{(2)},\nonumber\vspace{5pt}\\
		\label{Roeder eqs 6}
		\dfrac{\partial n_{\Omega}^{(3)}}{\partial t}+&v_\Omega^{(3)} \dfrac{\partial n_{\Omega}^{(3)}}{\partial a}=\omega^{(3)} n_A^{(3)}+r_{\text{inh}}n_\Omega^{(2)}+\\
		&+\left(\hspace{-2pt}-\dfrac{d v_{\Omega}^{(3)}}{d a}+\tau^{(3)}-\alpha^{(3)}-r_{\text{deg}}\hspace{-2pt}\right) n_\Omega^{(3)}\nonumber,
	\end{flalign}
\end{subequations}
where the super indexes $i$ represent the different cell populations as normal cells ($i=1$), imatinib-affected leukaemic cells ($i=2$) and non-affected leukaemic cells ($i=3$). Induced cell death is denoted by a constant $r_{\text{deg}}$, while the constant  $r_{\text{inh}}$ denotes the proliferation inhibition on the proliferating cells $n_{\Omega}^{(2)}$. The model in Eq. \eqref{Model Roeder} was proved to qualitatively and quantitatively reproduced the results of the agent-based approach for imatinib-treated patients in \cite{Roeder2006}. This was fitted to 894 peripheral blood samples, where the authors claimed that the therapeutic benefits of imatinib can, under certain circumstances, be accelerated by being combined with proliferation-stimulating treatment strategies. 

\cite{Kim2008a} described an extension of the \eqref{Model Roeder} model. This was done by considering the cycling cells $\Omega$ to be dependant, among other variables, on a counter $c(t)$, that indicates the position in the cell cycle, with a 49-hour cell cycle. An imatinib treatment was then incorporated into the model. The authors conclude that PDE formulation provided a more efficient way of simulating the dynamics of the disease. In fact, in simulations of imatinib treatment, the PDE and the discrete-time models diverged more, as in this case a continuous-time description of the disease dynamics may be more realistic than discrete-time models.
This latter model was later extended \cite{Clapp2014} by including feedback from cells and asymmetric division for stem cells and precursors. The general idea for this work was also to combine imatinib with a drug that induced cancer stem cells to cycle. Furthermore, the fact that many patients do relapse after being taken off imatinib motivates the study methods by which this therapy can be improved. \cite{Doumic-Jauffret2010} performed a stability analysis of the model in \cite{Kim2008a}, where the authors could set differences between AML and CML in terms of transition from stable equilibrium to unstable periodic behaviour.

\subsection{Whole body mathematical description of leukaemia and its treatment.}
\label{Section treatment general: several tissues}

\begin{figure}[ht]
	\centering	\includegraphics[width=0.8\textwidth]{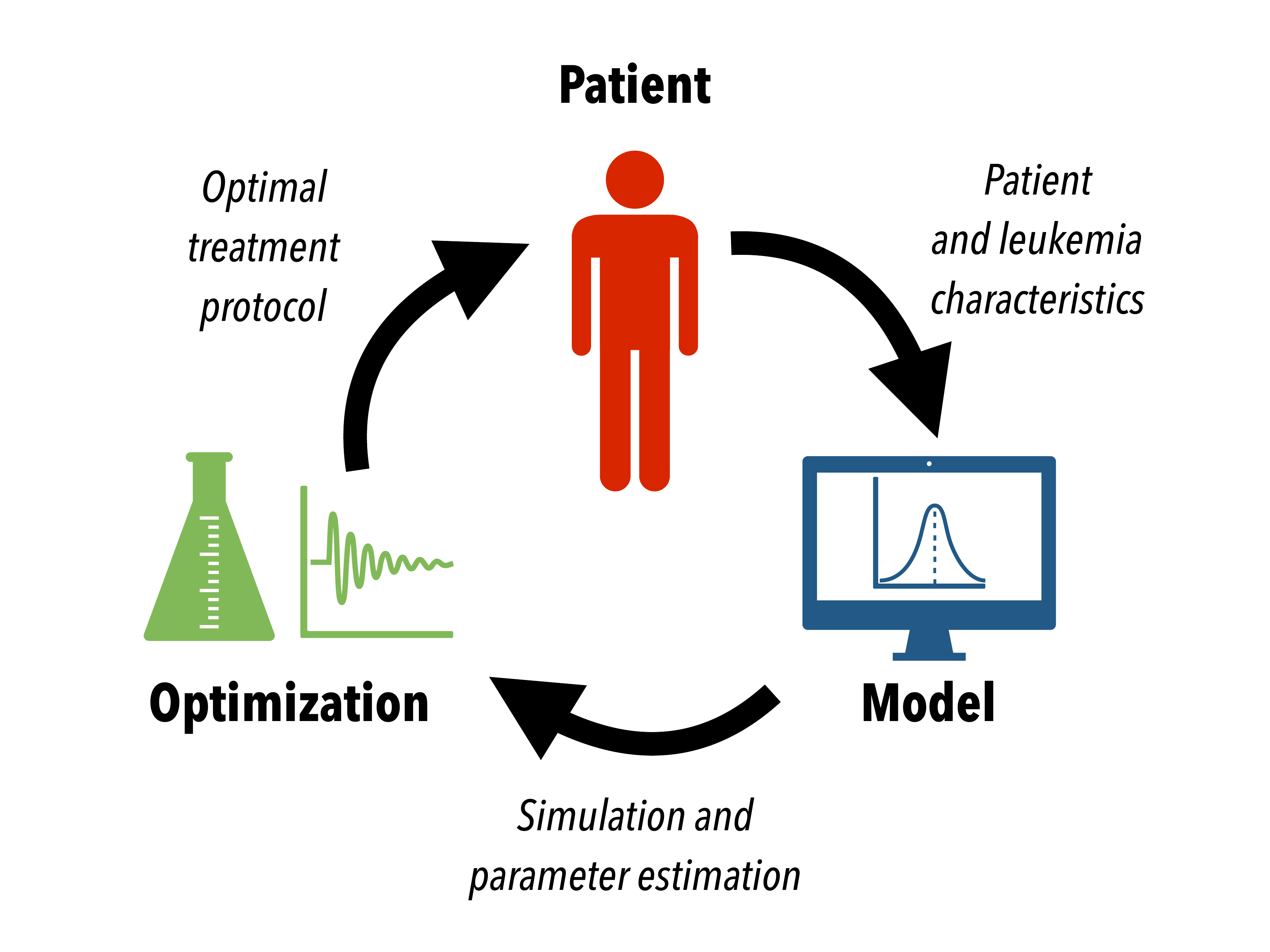}
	\caption{Schematic view of the use of mathematical models to help in patient treatment. Certain features are acquired from leukaemia, and specific ratios from each patient could be implemented in the mathematical model as parameters. After the simulation, several parameters might arise that could be advantageous for specific treatment protocol, optimised for the specific patient, who could benefit from the personalised drug. This cycle could be useful for following the disease in the patient. Fig. adapted from Ref. \cite{Pefani2014}.}
	\label{Fig modelizacion}
\end{figure}

Leukaemia treatment may affect blood flux in several tissues on the body. In order to understand the behaviour of these body parts during therapy, we set out a highly descriptive model of leukaemia, chemotherapy and blood flux throughout the entire body \cite{Pefani2014}. The inflow rate of drug $j$ is
\begin{subequations}
	\label{Pefani}
	\begin{flalign}
	\label{Pefani1}
	\text{inflow}_j=\dfrac{u_j}{duration_j},
	\end{flalign}
\end{subequations}
where $u_j$ is the drug dose over $duration_j$. This equation was then incorporated into the following equation, which models drug concentration in the blood $C_{B,j}$:
\addtocounter{equation}{-1}
\begin{subequations}
	\begin{flalign}\label{Pefani2}
	\setcounter{equation}{1}
	V_B\cdot \dfrac{dC_{B,j}}{dt}=&\sum_{i=H,Li,M,Le,K}Q_i\cdot C_{i,j}-Q_B\cdot C_{B,j}+\\
	&+\text{inflow}_j.\nonumber
	\end{flalign}
\end{subequations}
In this equation, $V_B$ is total patient blood volume, and $Q_i$ the blood flow in the organs $i$, such as heart ($H$), liver ($Li$), bone marrow ($M$), lean muscle ($Le$) and kidneys $K$, and so $C_{i,j}$ was the concentration of drug $j$ in the organs $i$, modelled as
\addtocounter{equation}{-1}
\begin{subequations}
	\begin{flalign}\label{Model all organs}
	\setcounter{equation}{2}
	V_i\cdot \dfrac{d C_{i,j}}{dt}=&Q_i\cdot C_{B,j}-Q_i\cdot C_{i,j}\\
	&-k_{k,j}\cdot C_{B,j}-k_{L,j}\cdot C_{i,j}\cdot V_{i,T},\nonumber
	\end{flalign}
\end{subequations}
for every organ $i$ and drug $j$, where  $k_{k,j}$ is the urine excretion rate, $k_{L,j}$ the elimination rate in the liver, and $V_{i,T}$ the volume of organ tissue where drug metabolism occurs. This model, along with many others, are useful in clinical terms, as it could provide guidance for optimising treatment for each patient in terms of their characteristics, as explained in Fig. \ref{Fig modelizacion}.

This pharmacokinetic model was reinforced by a pharmacodynamic model, which took into account the effect of the drug. Drug concentration at the location of the tumour, which for leukaemia would be the concentration of drug in the bone marrow ($C_{M,j}$), was considered for the $j$ effect of the drug as the function $\text{effect}_j$. It was included in the cell cycle as
\addtocounter{equation}{-1}
\begin{subequations}
	\begin{flalign}
	\setcounter{equation}{4}
	\dfrac{dP_y}{dt}=k_{y-1}\cdot P_{y-1}-k_y\cdot P_y-\text{effect}_j\cdot P_y,
	\end{flalign}
\end{subequations}
where $P_y$ was the cell population in phase $y$ ($G_1,S,G_2,M$) and $k_y$ the transition term from phase $y$ to $y+1$.

Although these equations are described in a general sense, for the specific case of chemotherapy cycles of intravenous ($IV$) daunorubicin ($DNR$) and cytarabine ($Ara-C$), typical drugs in leukaemia treatment, the reactions occurred at a subcutaneous level. That is, the drug is injected under the skin and not below muscle tissue. This drug and its subcutaneous effect have also been addressed in other studies, such as \cite{jost2019model1}, fitting data from 44 AML patients during consolidation therapy to a pharmacokinetic mathematical model, obtaining optimised treatment schedules. However, the authors of \cite{Pefani2014} considered, when simulating the subcutaneous effect of the therapy, that Eq. \eqref{Model all organs} could then be replaced by the following two:
\addtocounter{equation}{-1}
\begin{subequations}
	\begin{flalign}
	\setcounter{equation}{5}
	\dfrac{dS}{dt}=&\text{inflow}-k_a\cdot k_b \cdot S,\vspace{5pt}\\
	V _ { B } \cdot \frac { d C _ { B } } { d t } =& \sum _ { i =H , Li , M , L e , K } - Q _ { i}\cdot C_{i,j} - Q _ { B } \cdot C _ { B , j }+ \\
	&+ k _ { a } \cdot k _ { b } \cdot S,\nonumber
	\end{flalign}
\end{subequations}
where $S$ is the subcutaneous tissue drug delivery, $k_a$ the absorption delay and $k_b$ the drug bioavailability. However, the simulations performed were adapted for two acute myeloid leukaemia patients. Sensitivity analysis method was applied on the model to identify the most crucial parameters that control treatment outcome. The results clearly showed benefits from the use of optimisation as an advisory tool for treatment design.

The whole \eqref{Pefani} model was a clear example of the usefulness of mathematical models for therapy planning.

\section{Mathematical models of Acute Lymphoblastic Leukaemia treatments with cytotoxic drugs.}
\label{Section treatment ALL} 

\begin{figure}[ht]
	\centering\includegraphics[width=0.8\textwidth]{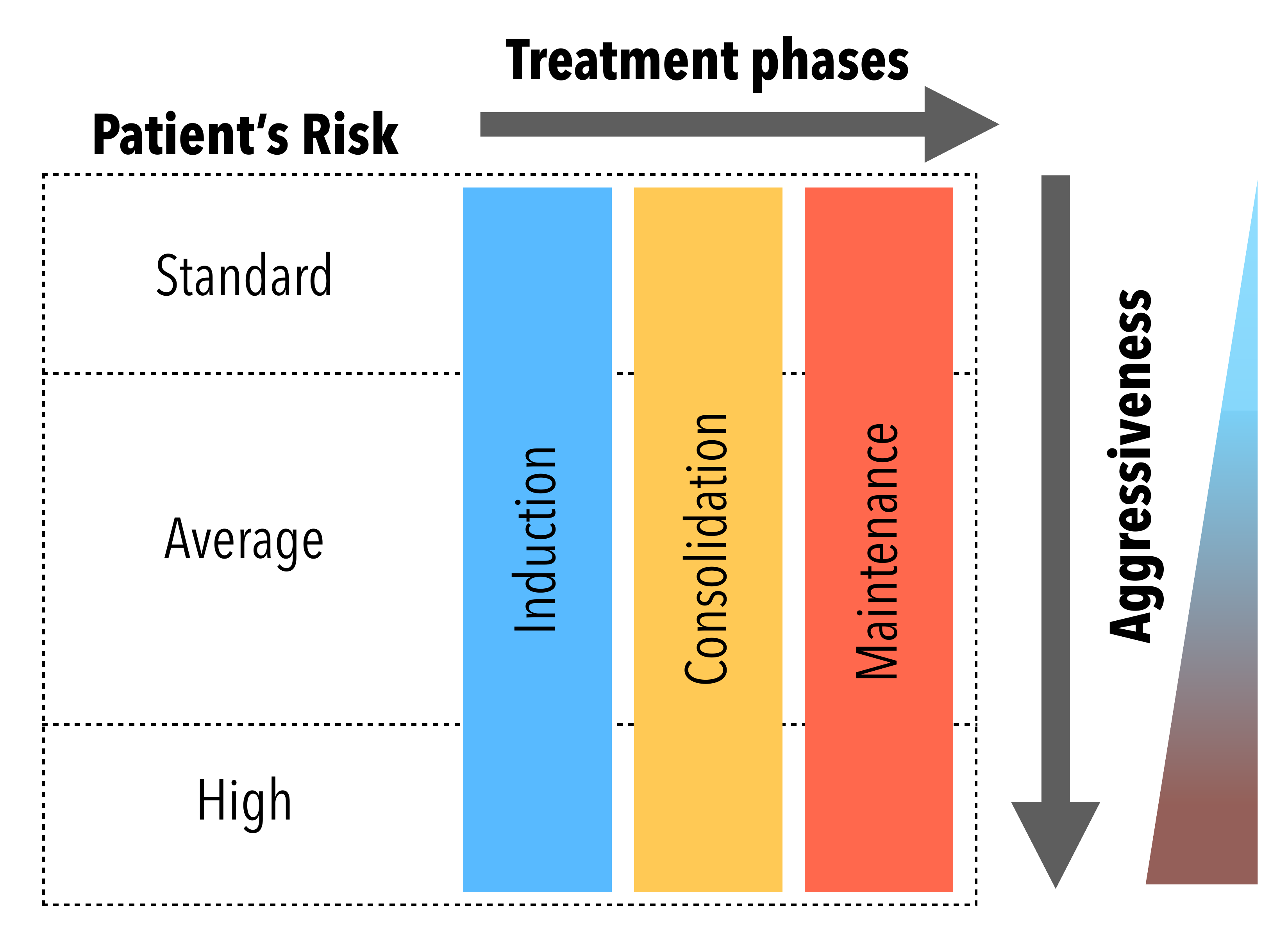}
	\caption{Stages of treatment administration for ALL, depending on the patient risk group.}
	\label{Treatment phases}
\end{figure}

The current standard treatment of acute lymphoblastic leukaemia involves different treatment stages: induction, consolidation, re-induction whenever needed, and maintenance \cite{Cooper2015}. The aggressiveness of treatments depends on the classification of patients into risk groups: standard, average or high (Fig. \ref{Treatment phases}).

The goal of the induction stage is to achieve a rapid reduction in tumour cell numbers. Next, the consolidation phase should ideally remove any trace of leukaemic cells in flow-cytometry or blood cell count studies. Re-induction is considered whenever leukaemic clones reappear early. The maintenance phase is administered when the first two phases are completed, and is intended to kill any possible remaining non-measurable quantities of cancer cells. Every phase includes specific treatments, the doses and timings of drugs depending on the patient’s risk group.

\begin{figure}[ht]
	\centering 	\includegraphics[width=0.8\textwidth]{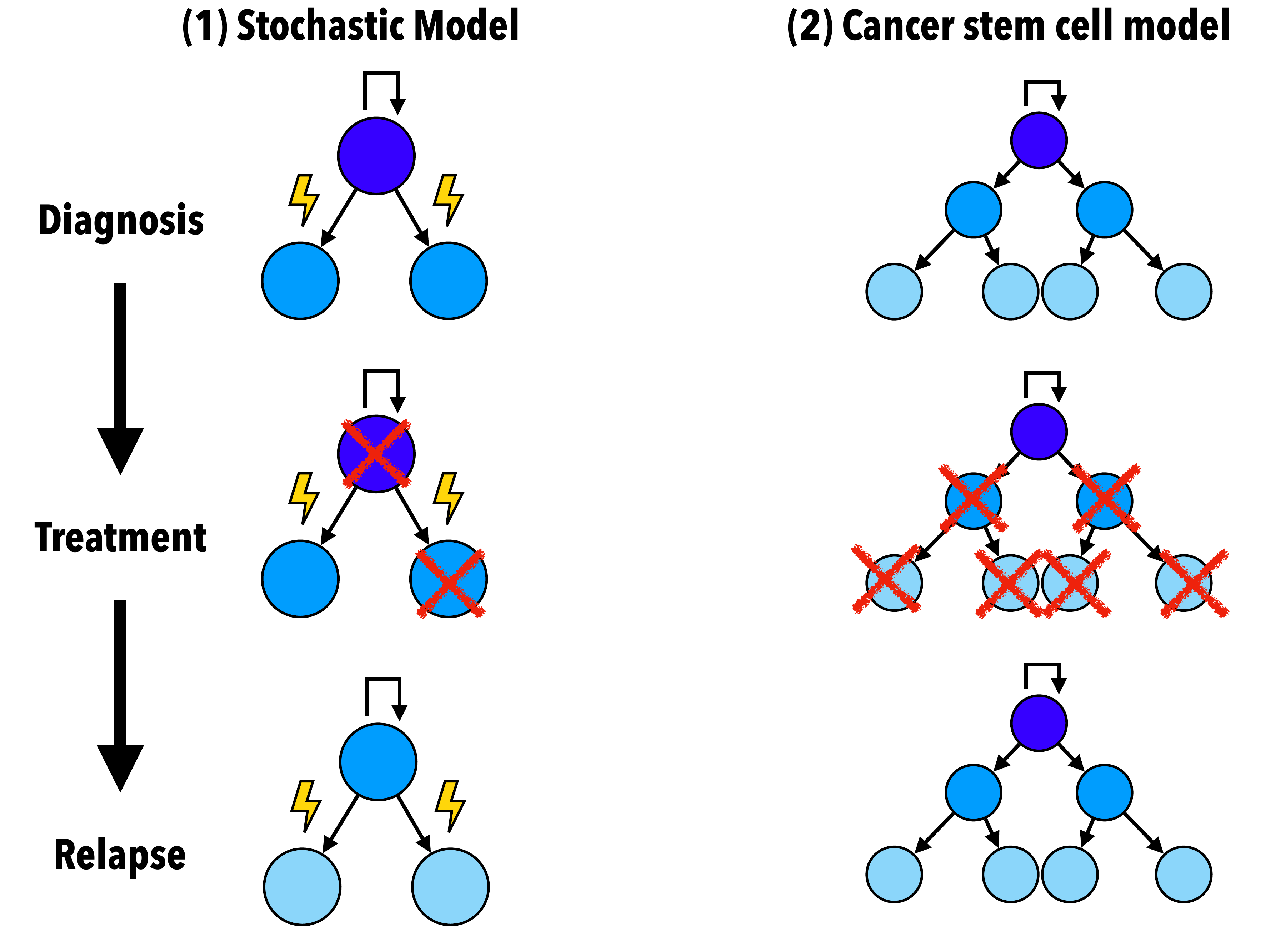} 
	\vspace{5pt}
	\caption{Different representations of tumour proliferation models through the effect of therapy. Figs. adapted from Ref. \cite{Lang2015}.} \label{Figure Therapy Stem Cells}
\end{figure}

Using one mathematical model or another to describe therapy may lead to a different understanding of how treatment affects cells in terms of relapse \cite{Lang2015}. For example, if relapse occurs and we consider a Cancer Stem Cell (CSC) model, a drug might not affect CSCs, or might only affect cells with specific mutations (in the genetic mutation model). This can be better seen in Fig. \ref{Figure Therapy Stem Cells}. 

In ALL, two drugs are used as part of these treament phases: 6-Mercaptopurine and Methotrexate.  Some mathematical models of their actions are now summarised. 

\subsection{Describing the effect of mercaptopurine.}
\label{Section ALL Mercaptopurine}

Mercaptopurine (6MP) is an antimetabolite antineoplastic agent with immunosuppressant properties. It interferes with nucleic acid synthesis by inhibiting purine metabolism and is used, usually in combination with other drugs, in the treatment of or in remission maintenance programmes for leukaemia.

A mathematical model of the effect of 6MP in leukaemia cells was described in \cite{Panetta2006}. In this model, the number of cells in the $G_0$/$G_1$-phases was denoted by $G$; $S$ in the $S$-phase, and $M$ in the $G_2$/$M$-phase. The suffixed variables $G_I$, $S_I$ and $M_I$ were the equivalent variables for the thioguanine (TGN) nucleotides, which were considered as the main active metabolites. That is, the most active molecules involved in the metabolic process. Apoptotic cells $A$, and non-viable cells $N$ (cells that are unable to live), were also included in the model. 

The equations for the viable phases of cells are
\begin{subequations}
	\label{Model Mercaptopurine}
	\begin{flalign}
		\label{Panetta TGNfree1}\dfrac{dG}{dt}&=-\alpha_S G+2\beta M,\vspace{5pt}\\
		\label{Panetta TGNfree2}\dfrac{dS}{dt}&=\alpha_S G-(\alpha_M +\gamma_1)S,\vspace{5pt}\\
		\label{Panetta TGNfree3}\dfrac{dM}{dt}&=(1-f)\alpha_M S - \beta M;
	\end{flalign}
\end{subequations}
while those for the cells with TGN incorporated are
\addtocounter{equation}{-1}
\begin{subequations}
	\addtocounter{equation}{3}
	\begin{flalign}
		\label{Panetta TGNwith1}\dfrac{dG_I}{dt}&=f\alpha_M S+\alpha_{MI}S_I-\beta_I M_I,\vspace{5pt}\\
		\label{Panetta TGNwith2}\dfrac{dS_I}{dt}&=-\alpha_{SI} G_I+2\beta_IM_I,\vspace{5pt}\\
		\label{Panetta TGNwith3}\dfrac{dM_I}{dt}&=\alpha_{SI}G_I - (\gamma_{MP}+\alpha_{MI})S_I.
	\end{flalign}
\end{subequations} 
Finally, the apoptotic and non-viable phases are modelled as  
\addtocounter{equation}{-1}
\begin{subequations}
	\addtocounter{equation}{6}
	\begin{flalign}
		\label{Panetta Apototic}\dfrac{dA}{dt}&=\gamma_1 S+\gamma_{MP} S_I -\gamma_2 A,\vspace{5pt}\\
		\label{Panetta Non viable}\dfrac{dN}{dt}&=\gamma_2A-\gamma_3 N.	\end{flalign}
\end{subequations}
\begin{figure}[ht]
	\centering\includegraphics[width=0.8\textwidth]{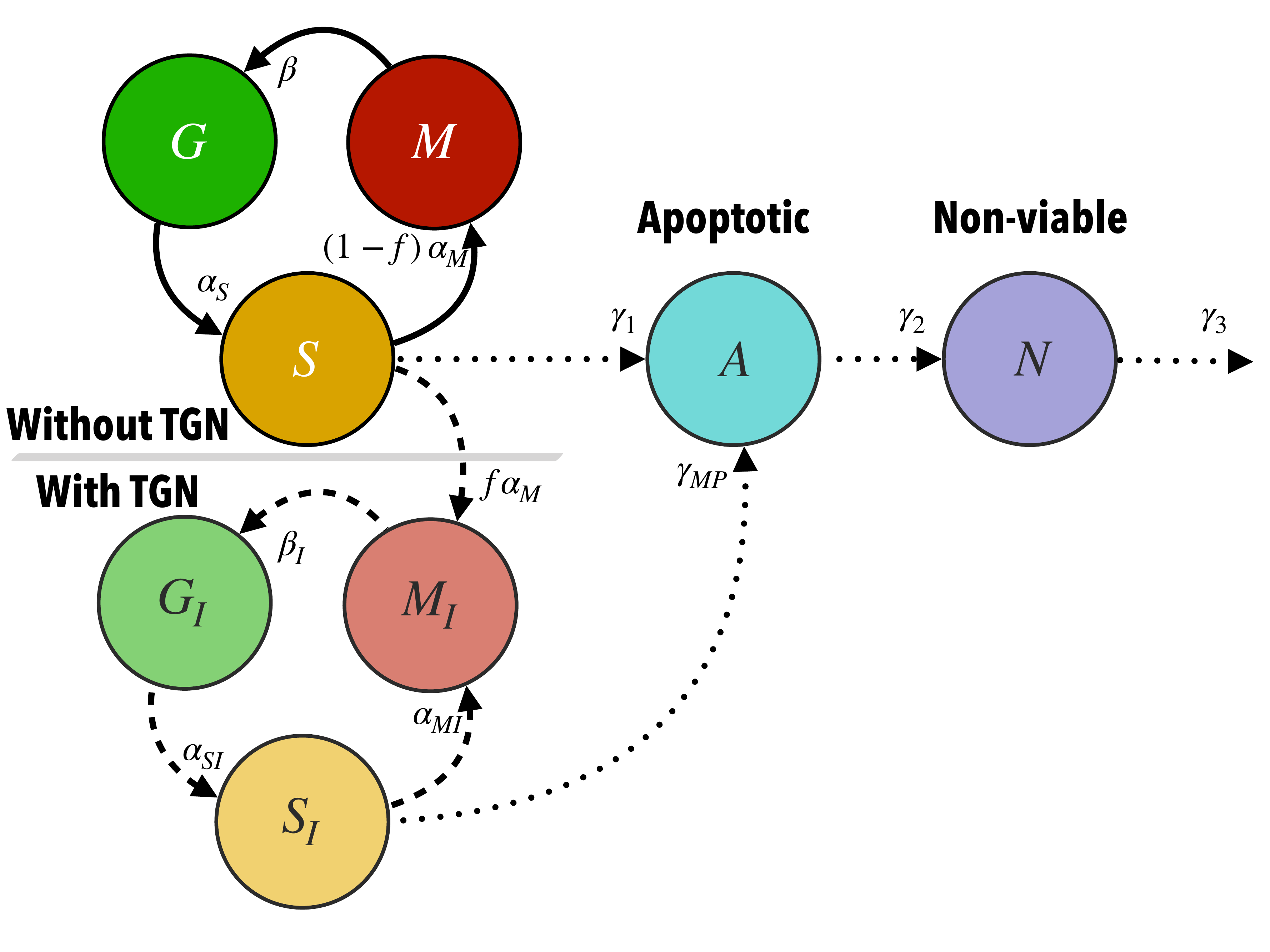}
	\caption{Diagram of the model \eqref{Model Mercaptopurine}. The different phases of the cell cycle are represented for cells with and without TGN incorporated into their DNA, before they reach the apoptotic and finally the non-viable state.}
	\label{FigMercaptopurine}
\end{figure}

The dynamics of the model are summarised in Fig. \ref{FigMercaptopurine}.  The model parameters describe the transition between phases, except for $f\in[0,1]$, which measures the fraction of cells continuing the cell cycle after TGNs were incorporated into the cell DNA. To estimate these parameters, the model was  fitted to data for different cell lines treated with MP.
The mathematical model provided a quantitative assessment to compare the cell cycle effects of MP in cell lines with varying degrees of MP resistance.

In a different study \cite{Jayachandran2014}, semi-mechanistic mathematical models were also designed and validated for MP metabolism, by studying red blood cell mean corpuscular volume (MCV) dynamics, a biomarker of treatment effectiveness and leukopenia, a major side effect related to very low percentages of leukocytes. The model was validated with real patient data obtained from literature and a local institution. Models were  individualised for each patient using nonlinear model-predictive control. The authors claimed that their approach could be implemented with routinely measured complete blood counts (CBC) and a few additional metabolite measurements. This would allow model-based individualised treatment, as opposed to a standard dose for all, and to prescribe an optimal dose for a desired outcome with minimum side-effects. 

\subsection{Mathematics of methotrexate treatments.}
\label{Section ALL Methotrexate}

Methotrexate (MTX) is an antimetabolite of the antifolate type. It is thought to affect cancer by inhibiting dihydrofolate reductase, an enzyme that participates in the tetrahydrofolate synthesis. This leads to an inhibitory effect on the synthesis of DNA, RNA, thymidylates, and proteins.

A first mathematical model of MTX effect in ALL was constructed in \cite{Panetta2423}.  
The authors based their approach on the fact that within cells, MTX is metabolised to more active methotrexate polyglutamates (MTXPG), and these polyglutamates are subsequently cleaved in lysosomes by glutamyl hydrolase (GGH). GGH acts as either an endopeptidase or an exopeptidase. To better define the in-vivo functions of GGH in human leukaemia cells, GGH activity was characterised with different MTXPG substrates in human T- and B-lineage leukaemia cell lines and primary cultures. Parameters estimated from fitting a series of hypothetical mathematical models to the data revealed that the experimental data were best fitted by a model where GGH simultaneously cleaved multiple glutamyl residues, with the highest activity on cleaving the outermost or two outermost residues from a polyglutamate chain. The model also revealed that GGH has a higher affinity for longer chain polyglutamates.

Further research led to the development of an improved model in \cite{Panetta2010}:
\begin{subequations}
	\label{Model Panetta MTX}	
	\begin{flalign}
		\dfrac{dMTX}{dt}=&-(k_e+k_{12})MTX+k_{21}MTX_p,\vspace{5pt}\\
		\dfrac{dMTX_p}{dt}=&k_{12}MTX-k_{21}MTX_p,\vspace{5pt}\\
		\dfrac{dMTXPG_1}{dt}=&\dfrac{V_{\max -in}MTX/V}{K_{m-in}+MTX/V}+k_pMTX/V-\\
		&-k_{eff}MTXPG_1+k_{GGH}MTXPG_{2-7}-\nonumber \vspace{5pt}\\
		&-\dfrac{V_{\max-FPGS}MTXPG_1}{K_{m-FPGS}+MTXPG_1},\vspace{5pt}\nonumber\\
		\dfrac{dMTXPG_{2-7}}{dt}=&\dfrac{V_{\max-FPGS}MTXPG_1}{K_{m-FPGS}+MTXPG_1}-\\
		&-k_{GGH}MTXPG_{2-7}.\nonumber
	\end{flalign}
\end{subequations}
This latter model simulated the concentration of MTXPG$_i$, where the subscripts denoted the number of glutamates attached to each MTX molecule. This provided new insights into the intracellular disposition of MTX in leukaemic cells and how it affects treatment efficacy. The variables $MTX$ and $MTX_p$ denoted the central and peripheral compartments of MTX. The parameters described: an elimination of plasma ($k_e$);  transition between peripheral and central compartments of MTX ($k_{12},k_{21}$); systemic volume ($V$); influx of MTX into the leukaemic blasts ($V_{\max-in}$, $K_{m-in}$);  first order influx and efflux ($k_p$ and $k_{eff}$, respectively); FPGS activity ($V_{\max-FPGS}$, $K_{m-FPGS}$) and $\gamma$-glutamyl hydrolase activity ($k_{GGH}$). Data from 791 plasma samples from 194 patients were used to validate the model. The study of the mathematical equations revealed that GGH activity had a higher affinity for longer chain polyglutamates and FPGS activity was higher in B-lineage ALL in comparison to T-lineage ALL.

Finally, \cite{Le2018} constructed a model involving a combination of several drugs, for chemotherapy-induced leukopenia in paediatric ALL patients. The model accounted for the action of both 6-MP and MTX and their cytotoxic metabolites 6-TGNc and MTXPGs during maintenance therapy. The equations were built on the basis of the previously discussed models  \cite{Panetta2010,Jayachandran2014}. The model predicted WBC counts for the available patient data surprisingly well, given the large variation of individual response patterns in the clinical data. The mathematical model and algorithmic procedure proposed could be used to guide personalised clinical decision support in childhood ALL maintenance therapy. Another model based on Refs. \cite{Panetta2010,Jayachandran2014} gave rise to a compartmental model in \cite{jost2019model2}, including pharmacokinetics and a myelosuppression model for ALL, considering both 6-MP and MTX. The model was cross-validated with data from 116 patients, and simulations of different treatment protocols were performed to exploit the optimal effect of maintenance therapy on survival.	

\section{Modelling immune response and immunotherapy in leukaemias.}
\label{Section treatment general: interactions}

\subsection{Immune response mathematical models.}
Immunotherapy is a type of therapy that stimulates cells within the immune system in order to help the body fight against cancer or infections. Interactions between cells are key in understanding processes such as, for example, proliferation or resource competition between cells.  The immune system is one way in which the body may influence external agents and a greater understanding of it could be useful in fighting leukaemia.

An extension of the model already described, \eqref{Model Michor}, was introduced in \cite{Clapp2015BCR}, where the CML populations were distributed as stem cells ($y_0$), progenitor ($y_1$) and mature leukaemic cells ($y_2$). In this study, the concentration of immune cells was also included and denoted as $z$. The authors designed a mathematical model integrating CML and an autologous immune response to the patients' data by considering the following system
\begin{subequations}
	\begin{flalign}
		\label{Model Clapp y0}
		\dfrac{dy_0}{dt}&=b_{1} y_{1}-a_{0} y_{0}-\dfrac{\mu y_{0} z}{1+\varepsilon y_{3}^{2}},\vspace{5pt}\\ 
		\label{Model Clapp y1}
		\dfrac{dy_1}{dt}&=a_{0} y_{0}-b_{1} y_{1}+r y_{1}\left(1-\dfrac{y_{1}}{K}\right)-d_{1} y_{1}-\dfrac{\mu y_{1} z}{1+\varepsilon y_{3}^{2}},\vspace{5pt}\\
		\label{Model Clapp y2}
		\dfrac{dy_2}{dt}&=a_{1} y_{1}-d_{2} y_{2}-\dfrac{\mu y_{2} z}{1+\varepsilon y_{3}^{2}}, \vspace{5pt}\\
		\label{Model Clapp y3}
		\dfrac{dy_3}{dt}&=a_{2} y_{2}-d_{3} y_{3}-\dfrac{\mu y_{3} z}{1+\varepsilon y_{3}^{2}}, \vspace{5pt}\\ 
		\label{Model Clapp z}
		\dfrac{dz}{dt}&=s_{z}-d_{z} z+\dfrac{\alpha y_{3} z}{1+\varepsilon y_{3}^{2}},
	\end{flalign}
\end{subequations}
where $a_0,b_1$ represents transition terms; $d_z$ and $d_i$, for each cell type $i=1,2,3$, denotes cell death; and a logistic growth for progenitor cells $y_1$ was included, with a reproduction rate $r$. The immune system action rate $\mu$ was included in the mass action term ``$\mu\, y_i\, z$'' in the last term of the leukaemic population equations from Eq. \eqref{Model Clapp y0} to Eq. \eqref{Model Clapp y3}. The proliferation of the immune system pool included a constant factor $s_z$ and was activated by mature leukaemia cells with the term ``$\alpha\, y_3\, z$''  in Eq.  \eqref{Model Clapp z}. These latter terms included an inhibition of the immune cells expansion, as they were divided by ``$1+\epsilon y_3^2$'', where $\epsilon$ was the strength of the immunosuppression. This model included data from patients treated with imatinib, and their BCR-ABL transcripts, related to leukaemia diagnosis. The authors considered that variations in BCR–ABL  transcripts during imatinib therapy may represent a signature of the patient's individual autologous immune response. The use of immunotherapy was then considered to be a useful complement to the usual treatment, playing a significant role in eliminating the residual leukaemic burden.

A general mathematical model for tumour immune resistance and drug therapy was proposed in \cite{Pillis2001}. By including tumour cells, immune cells, host cells and drug interaction, an optimal control problem was constructed. This would provide a basis for the study of leukaemia immune cell interaction, shedding some light on the modelling for B leukaemia. For B-cells, fundamental in both acute and chronic lymphocytic leukaemia diagnosis, a more extensive model was presented in \cite{Nanda2013}, including four different cell populations in the peripheral blood of humans: B cells, able to bind to antigens which will initiate antibody responses; NK cells, critical to the immune system; cytotoxic T cells, able to kill cancer cells; and helper T cells, which may help other immune cells by releasing T cell cytokines. This model was considered a tool that may shed light on factors affecting the course of disease progression in patients, and focused on sensitivity analysis for parameters and bifurcation analysis.  Based on \cite{Pillis2001}, an immunotherapy approach was considered in \cite{Rodrigues2019} by developing a model focused on B and T lymphocytes and their relation with a chemotherapeutic agent. The ODE system for this model was the following
\begin{subequations}
	\begin{flalign}
		\label{Rodrigues1}
		\dfrac{d N}{d t}&=r N\left(1-\dfrac{N}{k}\right)-c_{1} N I-\dfrac{\mu N Q}{a+Q}, \\ 
		\label{Rodrigues2}
		\dfrac{d I}{d t}&=s(t)+s_{0}-d I+\dfrac{\rho N I}{\gamma+N}-c_{2} N I-\dfrac{\delta I Q}{b+Q} ,\\
		\label{Rodrigues3}
		\dfrac{d Q}{d t}&=q(t)-\lambda Q,\end{flalign}
\end{subequations}
where $N=N(t)$ represented the neoplastic B lymphocytes, $I=I(t)$ the healthy T lymphocytes (this is, the immune cells), and $Q=Q(t)$ the amount of a chemotherapeutic agent in the bloodstream. In Eq. \eqref{Rodrigues1} $N$ follows a logistic growth with a proliferation rate $r$, and dies due to both interaction with immune cells at a rate $c_1$ and with the chemotherapeutic agent at a rate $\mu$. Immune cells in Eq. \eqref{Rodrigues2} have a constant source $s_0$ and die naturally at a constant rate $d$ and also due to interaction with cancer cells at a rate $c_2$, and with drugs at a rate $\delta$. However, there is a production rate $\rho$ of immune cells stimulated by cancer cells. Both $N$ and $I$ have Michaelis-Menten terms with rates $a$, $\gamma$ and $b$. For the case of the chemotherapeutic agent $Q$ in Eq. \eqref{Rodrigues3}, $\lambda$ is considered as the washout rate of a given cycle-nonspecific chemotherapeutic drug with $\lambda=\ln(2)/t_{\frac{1}{2}}$, where $t_{\frac{1}{2}}$ is the drug elimination half-life. Finally, the functions $s(t)$ and $q(t)$ are source terms, which can be considered to be constants. These parameters were all taken from the literature and claimed to simulate CLL behaviour. This model reinforces the option of combining treatments such as chemo- and immunotherapy, where the first may decrease cells to a point where immune cells may act. 

A model for AML was considered in \cite{Nishiyama2017} by including the role of leukaemic blast cells $(L)$, mature regulatory T cells $(T_\text{reg})$ and mature effector T cells $(T_\text{eff})$, this last also including cytotoxic T lymphocytes and Natural Killers. The aim of including such cells was to create an activated immune cell infusion with selective $T_\text{reg}$ depletion.  This was done by converting the intracellular interaction into a model, as the following system:
\begin{subequations}
	\begin{flalign}\frac{d[L]}{d t}&=a_{L}\left(\frac{k_{1}^{p}}{k_{1}^{p}+\left[T_\text{eff}\right]^{p}}\right)-d_{L}[L], \\ 
		\frac{d\left[T_\text{eff}\right]}{d t}&=a_{T_\text{eff}}\left(\frac{k_{2}^{p}}{k_{2}^{p}+\left[r_\text{reg}\right]^{p}}\right)-d_{T_\text{eff}}\left[T_\text{eff}\right], \\
		\frac{d\left[T_\text{reg}\right]}{d t}&=a_{T_\text{reg}}\left(\frac{[L]^{p}}{k_{3}^{p}+[L]^{p}}\right)-d_{T_\text{reg}}\left[T_\text{reg}\right],
	\end{flalign}
\end{subequations}
where $a_L$, $a_{T_\text{eff}}$, $a_{T_\text{reg}}$ represented influx rates, and $d_L$, $d_{T_\text{eff}}$, $d_{T_\text{reg}}$ the decay rates.  Intercellular interactions were modelled as Hill functions with threshold constants ($k_1$, $k_2$, $k_3$) with strength $p$. Two existing steady states were found for this model in \cite{Nishiyama2017}, corresponding to leukaemia diagnosis or relapse, and to complete remission. The authors considered that the model explained the influence of the duration of complete remission on the survival of patients with AML after allogeneic stem cell transplantation. In \cite{Nishiyama2018}, simulations were run for this model by performing Monte Carlo simulation of trajectories in the phase plane, and generated relapse-free survival curves, which were then compared with clinical data. This provided  valuable information for the future design of immunotherapy in AML.

\subsection{Including interleukins in mathematical models.}
Interleukins (ILs) are a group of cytokines first seen to be expressed by white blood cells (leukocytes). The immune system depends on interleukins as these signals between cells are useful for acting against several pathogens. 

The interaction between the actively responding effector cells $E=E(t)$, tumour cells ($T=T(t)$) and the concentration of the cytokine IL-2 ($I_L=I_L(t)$)  was the basis for the latter study, influenced by \cite{Kirschner1998}. The reason behind the modelling of this cytokine is due to the fact that IL-2 might boost the immune system to fight tumours. This was described via the following system:
\begin{subequations}
	\begin{flalign}
		\dfrac{dE}{dt}&=cT-\mu_2E+\dfrac{p_1\,E\,I_L}{g_1+I_L}+s_1,\vspace{5pt}\\
		\dfrac{dT}{dt}&=r_2(T)-\dfrac{a\,E\,T}{g_2+T},\vspace{5pt}\\
		\dfrac{dI_L}{dt}&=\dfrac{p_2\,E\,T}{g_3+T}-\mu_3 I_L+s_2.
	\end{flalign}
\end{subequations}
In this model, $c$ was antigenicity or ability to provoke an immune response, $\frac{1}{\mu_2}$ was the average natural lifespan, $a$ the loss of tumour cells by interaction, $\mu_3$ the degraded rate of IL-2,  and $s_1$, $s_2$ were treatment terms. The fraction terms were of the Michaelis-Mentis form, to indicate saturation effects. The function $r_2(T)$ could be described as a constant for linear growth, or with limiting-growth as logistic or Gompertz terms. With this model, the authors concluded that with only IL-2 treatment, the immune system might not be enough to clear tumours. These and other models were reviewed in \cite{Talkington2017} in terms of equilibrium points, considering T lymphocytes and their interaction with other cells, and it was found that there are two stable equilibrium points, one where there is no tumour, and the other where there is a large one.

Interaction between cells via interleukins was also studied in \cite{Cappuccio2006}, as IL-21 is being developed as an immunotherapeutic cancer drug. Its effect has been studied in relation to Natural Killer (NK) cells, and CD8\textsuperscript{+} T-cells, which have the ability to make cytokines, with the model
\begin{subequations}
	\label{Cap}
	\begin{flalign}
		\label{Cap A}
		\dfrac{du}{dt}&=\text {input} - \mu _ { 1 } u,\\
		\label{Cap B}
		\dfrac{dx}{dt}&=r _ { 1 } x \left( 1 - \frac { x } { h _ { 1 } ( u ) } \right),\\
		\label{Cap C}
		\dfrac{dy}{dt}&=r _ { 2 } y \left( 1 - \frac { y } { h _ { 2 } ( m ) } \right),\\
		\label{Cap D}
		\dfrac{dm}{dt}&=a u - \mu _ { 2 } m,\\ 
		\label{Cap E}
		\dfrac{dp}{dt}&=\frac { b _ { 1 } u } { b _ { 2 } + u } - \mu _ { 3 } p,\\ 
		\label{Cap F}
		\dfrac{dn}{dt}&=g ( n ) - k _ { 1 } p x n - k _ { 2 } p y m, 
	\end{flalign}
\end{subequations}
where Eq. \eqref{Cap A} represented the concentration of IL-21, Eq. \eqref{Cap B} the concentration of NK in the spleen, Eq. \eqref{Cap C} the antitumour CD8\textsuperscript{+} T-cells in the lymph, Eq. \eqref{Cap D} a facilitating T-cell memory factor useful for expressing the recognition of foreign invaders for memory T-cells, Eq. \eqref{Cap E}  a cytotoxic protein affecting tumour lysis, and finally tumour mass at any time was represented by Eq. \eqref{Cap F}. The functions involved were defined in the monotonic decreasing function
\addtocounter{equation}{-1}
\begin{subequations}
	\begin{flalign}
		\setcounter{equation}{6}
		h _ { 1 } ( u ) = \frac { p _ { 1 } u + p _ { 2 } } { u + q _ { 1 } },
	\end{flalign}
\end{subequations}
the function of the memory factor $m$
\addtocounter{equation}{-1}
\begin{subequations}
	\begin{flalign}
		\setcounter{equation}{7}
		h _ { 2 } ( m ) = h _ { 2 } ( 0 ) + \frac { \sigma m } { 1 + \frac { m } { D } },
	\end{flalign}
\end{subequations}
and $g(n)$ the dynamics of tumour cell number, which is constructed separately for each tumour type according to the observed growth curves. Parameters were estimated in terms of certain values from the literature, so that simulations were run to show IL-21 as a promising antitumour therapeutic. For more immunotherapeutic approaches towards cancer modelling, we highlight the work in \cite{Preziosi2003}, where some general aspects of cancer were also reviewed, including diffusion, angiogenesis and invasion.

Finally, for the case of immune response to leukaemia, other studies have been undertaken, though not specially in the form of an ODE or PDE system. Some numerical simulations were run in \cite{Kolev2005} by proposing an integro-differential equation model. This study proposed a new possibility for defining the activation states for cancer, cytotoxic T and T helper cells. Using these definitions, the authors suggested that it would be easier to organise experiments suitable for measuring cell states. They also claimed that cell-mediated immunity is one of the most crucial components of antitumour immunity. Immune T-cells were studied in \cite{Chrobak2012} in terms of a stochastic model from which was derived a Fokker-Planck equation. Stability analysis and behaviour of the solutions of the model led to the conclusion that more accurate simulations of cancer genesis and treatment were needed. Lastly, in \cite{Saadatpour2011}, cytotoxic T cells were dynamically and structurally analysed in terms of a Boolean network model for T cell large granular lymphocyte leukaemia. Nineteen potential therapeutic targets were found, and these were versatile enough to be applicable to a wide variety of signals and regulatory networks related to diseases.   

\subsection{Novel therapies for leukaemia models: CAR-T cells.}
Immunotherapy based on chimeric antigen receptor T  (CAR-T) cells has been especially successful in patients who did not respond to the usual types of chemotherapy.  This technique is based on the patient's own T-cells, which are extracted from them, genetically modified and reinfused.  This modification allows T-cells to kill tumour cells in a more effective way than the usual chemotherapies.

\begin{figure*}[!ht]
	\centering
	\includegraphics[width=\textwidth]{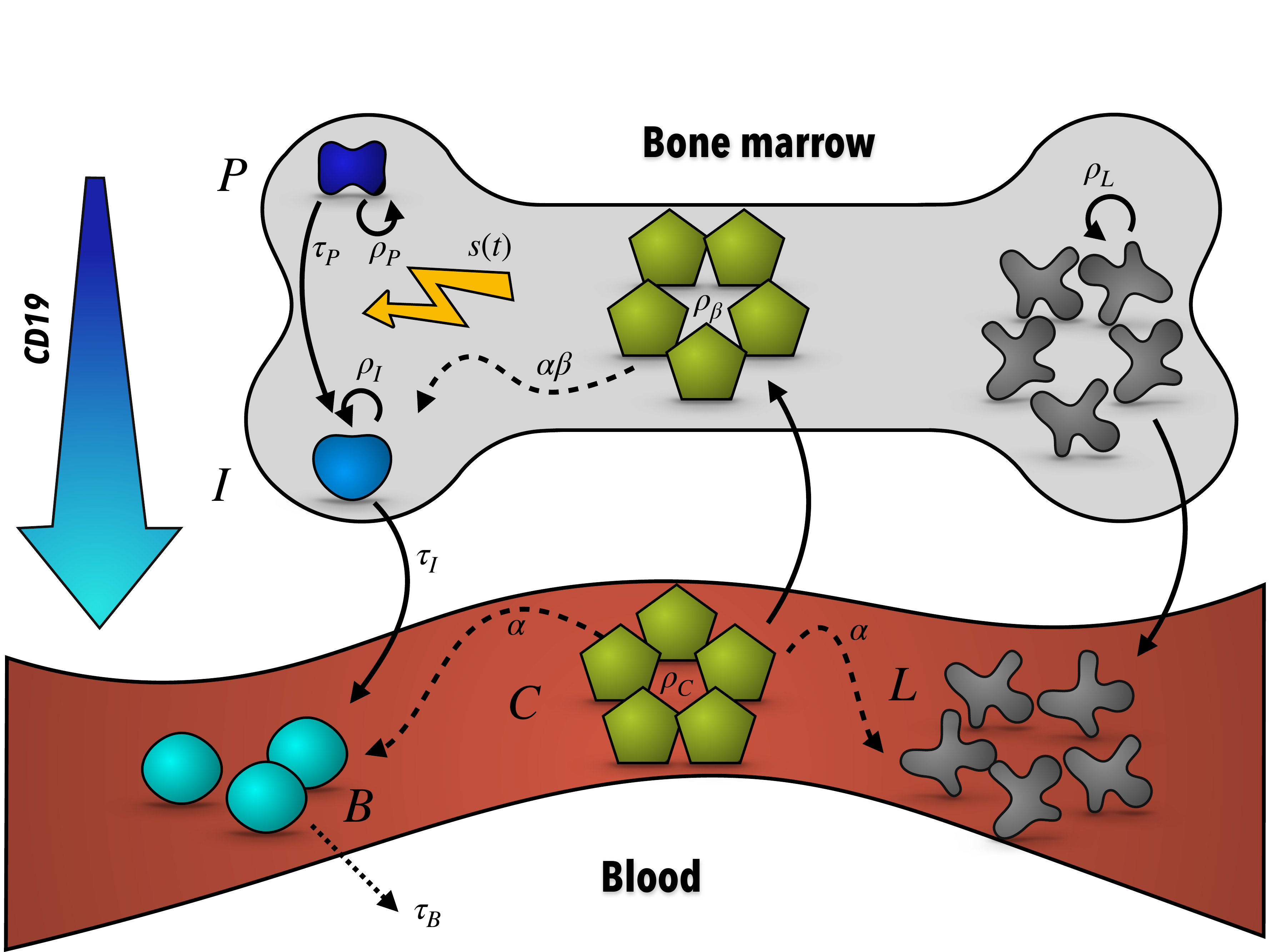}
	\caption{Illustration of the dynamics in model \eqref{leon model CAR-T}. B cells (in blue) develop in the bone marrow, arising from progenitor CD19\textsuperscript{-} cells ($P$), then turning, with rate $\tau_P$, into CD19\textsuperscript{+} cells $I$ and reaching, with rate $\tau_I$ a mature stage of healthy B cells $B$, finally dying after a time $\tau_B$. During this process, a signalling effect $s(t)$ affects the proliferation rates of the early stages $\rho_P$ and $\rho_I$. Leukaemic cells $L$ develop in the bone marrow with rate $\rho_L$, invading this tissue as well as the blood compartment. CAR-T cells $C$ attack mature B cells and leukaemic cells with rate $\alpha$, also inducing growth, with rate $\rho_C$. In the bone marrow, they also attack CD19\textsuperscript{+} cells $I$, with a lower rate $\alpha \beta$. This interaction induces growth with rate $\rho_\beta$. Solid lines represent cell growth and change between compartments. The dotted line represents the natural death of the healthy B cells. Dashed lines represent cell death due to CAR-T cell interaction.}
	\label{figure CAR-T leon}
\end{figure*}

We have designed a general model for CAR-T cells in \cite{perez2020car} considering several cell compartments. Firstly, for T-cell leukaemia, the number of CAR-T cells was denoted by $C$, leukaemic T cells by $L$, and normal T-cells by $T$. The dynamics of the model were as follows
\begin{subequations}
	\begin{flalign}
		\frac{d C}{d t} &=\rho_{C}(T+L+C) C-\frac{1}{\tau_{C}} C-\alpha C^{2}+\rho_{I} C ,\\
		\frac{d L}{d t} &=\rho_{L} L-\alpha L C ,\\
		\frac{d T}{d t} &=g(T, L, C)-\alpha T C.
	\end{flalign}
\end{subequations}
The parameter $\rho_L$ represents leukaemic proliferation rate, while $\rho_C$ represents stimulation of CAR-T cell mitosis after encounters with target cells; $\tau_C$  is the finite lifespan of CAR-T cells; the parameter $\alpha$ represents death due to encounters with CAR-T cells; parameter $\rho_I$ is the external cytokine signal strength used for division of CAR-T cells; finally, the function $g(T,L,C)$ denotes the rate of production of normal T cells, assumed to contribute only at a minimal residual level. The stability analysis of the cell dynamics leads to several conclusions: firstly, CAR-T cells allow for control of T-cell leukaemia in the presence of fratricide; secondly, the initial number  of CAR-T cells injected, as well as re-injections, does not affect the outcome of therapy, while higher mitotic stimulation rates do; lastly, tumour proliferation rates have an impact on relapse time. A second, similar model was constructed for B cells, in \cite{leon2020car}, where CAR-T and now leukaemic B cells where again denoted as $C$ and $L$, but the inclusion of mature healthy B cells $B$,  CD19\textsuperscript{-} B cells $P$, and CD19\textsuperscript{+} cells $I$ was considered. The initial autonomous system of differential equations was
\begin{subequations}
	\label{leon model CAR-T}
	\begin{flalign}
		\frac{d C}{d t} &=\rho_{C}(L+B) C+\rho_{\beta} I C-\frac{1}{\tau_{C}} C, \\
		\frac{d L}{d t} &=\rho_{L} L-\alpha L C ,\\
		\frac{d B}{d t} &=\frac{1}{\tau_{I}} I-\alpha B C-\frac{1}{\tau_{B}} B,\\
		\frac{d P}{d t} &=\rho_{P}\left(2 a_{P} s(t)-1\right) P-\frac{1}{\tau_{P}} P ,\\
		\frac{d I}{d t} &=\rho_{I}\left(2 a_{I} s(t)-1\right) I-\frac{1}{\tau_{I}} I+\frac{1}{\tau_{P}} P-\alpha \beta I C,
	\end{flalign}
\end{subequations}
where parameters $\rho_C,\tau_C,\rho_L$ and $\alpha$ were the same as the considered in the previous model. Parameter $\rho_B= \beta\rho_C$, where $0 <\beta < 1$, accounts for the fact that represented B cells are located mostly in the bone marrow and encounters with CAR-T cells will be less frequent. Parameters $\rho_P$ and $\rho_I$ represent  growth rates for $P$ and $I$ cells, while $\tau_I,\tau_B$ and $\tau_P$ represent the finite lifespan of $I,B$ and $P$ cells respectively.  A signalling function $s(t)=1/[1+k_s(P+I)]$, with $k_s >0$ was constructed as in \cite{Marciniak-Czochra2009}, also including the asymmetric division rates $a_P$ and $a_I$ for $P$ and $I$. This general model is reduced, in order to understand the dynamics of the expansion of CAR-T cells and their effect on the healthy B and leukaemic cells, neglecting the contribution of the haematopoietic compartments. Parameters are estimated from the literature and the main conclusion obtained is that not only does CAR-T cell persistence depend on T-cell mean lifetime, but also that reinjection may allow the severity of relapse to be controlled. The dynamics of the model from Eq. \eqref{leon model CAR-T} are summarised in Fig. \ref{figure CAR-T leon}.

A general model taken from the literature and applied to CAR-T cells is set out in \cite{Khatun2020}.  The authors denote  $s=s(t)$ as the population of susceptible blood cells, $i=i(t)$ as the population of infected blood cells, $c=c(t)$ as the population of leukaemic cells (abnormal cells), and $w=w(t)$ as the population of white blood cells or immune cells. The dynamics are modelled as
\begin{subequations}\begin{flalign}{}
		\frac{d s}{d t}&=A-a_{0} s-\beta s c ,\\
		\frac{d i}{d t}&=\beta s c-\beta_{0} i-\beta_{1} c i ,\\
		\frac{d c}{d t}&=k-k_{0} c-k_{1} c w ,\\
		\frac{d w}{d t}&=B+b c-b_{0} w-b_{1} w c,
	\end{flalign}
\end{subequations}
where $a_0, \beta_0, k_0,$ and $b_0$ are the natural death rate of susceptible blood cells, infected cells, cancer cells, and immune cells, respectively; for susceptible cells, $A$ is the recruitment rate and $\beta$ is the loss rate of susceptible blood cells due to infection; $\beta_1$ is the decay rate parameter of infected cells; $k$ is the constant recruitment rate of cancer cells, while $k_1$ and $b_1$ are the loss rates of cancer and immune cells due to interaction; finally, parameter $B$ is considered as the external re-infusion rate of immune cells (CAR-T). This model was studied in terms of stability, and it was observed that the external re-infusion of immune cells by adoptive T-cell therapy reduces the concentration of cancer cells and infected cells in the blood.

With the success of T-cell-engaging immunotherapeutic agents, there has been growing interest in the so-called cytokine release syndrome (CRS), as it represents one of the most frequent serious adverse effects of these therapies. CRS is a systemic inflammatory response that can be caused by a variety of factors, such as infections and certain drugs. A more specific model that included the action of cytokines was studied by considering Tisagenlecleucel, a personalised cellular therapy of CAR-T cells for B-cell ALLs, associated with a high remission rate. It was modelled in \cite{mostolizadeh2018mathematical}  by considering the interaction of a CAR-T cell population $c_T=c_T(t)$  with B-cell leukaemic population $l=l(t)$, as well as with healthy B cells $h=h(t)$, both marked with CD19, a characteristic of B lymphocytes. Other circulating lymphocytes were denoted as $c=c(t)$, while the number of cytokines, key to understanding inflammatory processes, was generally considered as $s=s(t)$. The dynamics of the model were as follows:
\begin{subequations}
	\begin{flalign}{}
		\label{Tisganmarciniak1}
		\frac{dc_{T}}{dt}&=d_{1} c_{T}-d_{2} c_{T}-\alpha_{1} c_{T} l-\beta_{1} c_{T} h, \\
		\label{Tisganmarciniak2}
		\frac{di}{dt}&=k l-\alpha_{2} c_{T} l, \\
		\label{Tisganmarciniak3}
		\frac{dh}{dt}&=a h(1-b h)-d_{3} h-\beta_{2} c_{T} h ,\\
		\label{Tisganmarciniak4}
		\frac{dc}{dt}&=\lambda-\sigma c+\alpha_{3} \frac{c_{T} c}{\beta_{3}+c_{T}} ,\\
		\label{Tisganmarciniak5}
		\frac{ds}{dt}&=\alpha_{4}-\beta_{4} s+d_{4}\left(\frac{c_{T}}{c_{T}+m}\right),
	\end{flalign}
\end{subequations}
where Eq. \eqref{Tisganmarciniak1} represented the dynamics of CAR-T cells with growth rate $d_1$ and natural death rate $d_2$, while $\alpha_1$ and $\beta_1$ were cell death given by interaction with leukaemic and healthy cells, respectively. Eq. \eqref{Tisganmarciniak2} includes a growth rate of leukaemic cells $k$ and a cell death $\alpha_2$ by interaction with $c_T$. Eq. \eqref{Tisganmarciniak3} described a logistic growth of healthy cells with rates $a$ and $b$, as well as a natural death rate $d_3$  and death $\beta_2$ due to interaction with $c_T$. Circulating lymphocyte dynamics were considered in Eq. \eqref{Tisganmarciniak4} to have a constant input $\lambda$, death rate $\sigma$ and growth dependant on $c_T$, attenuated via a Hill function with constants $\alpha_3$ and $\beta_3$. Finally, for Eq. \eqref{Tisganmarciniak5}, cytokines were secreted at a maximum rate $\alpha_4$ and altered by a negative feedback mechanism corresponding to the term $-\beta_{4}s$. Furthermore, the stimulation of CAR-T cells increased the levels of cytokines with rate $d_4$ and a constant $m$ from the correspondent Hill function. Optimal control theory was applied for this model, controlling the injection of CAR-T cells and cytokines, to finally minimise the level of cancer cells and to keep healthy cells above a desired level.

Effector T cells are a group of cells including several T-cell types that actively respond to a stimulus. Following an infection, memory T cells are antigen-specific T cells that remain in the long term. This distinction is considered to help understand the dynamics of CAR-T cells in several models. For instance, a general description of Tisagenlecleucel was performed in \cite{Stein2019}, where data from 91 paediatric and young adult B-ALL patients were used for the analysis. The model describes the expansion of CAR-T cells up to a time $T_{\max}$, and then two phases: a first contraction phase, with rapid decline; and a second persistence phase, declining more gradually. This was represented by a dynamic system considering effector $E$ and memory CAR-T cells $M$, as
\begin{subequations}
	\begin{flalign}
		\frac{d E}{d t}&=\rho \cdot F(t) \cdot E, \quad \text{for }T\leq T_{\max},\\
		\frac{d E}{d t}&=-\alpha \cdot E ,\quad\text{for }T> T_{\max},\\
		\frac{d M}{d t}&=k \cdot E-\beta \cdot M,\quad\text{for }T> T_{\max},
	\end{flalign}
\end{subequations}
and $M=0$,  for $T\leq T_{\max}$. After $T_{\max}$, effector cells rapidly decline at a rate $\alpha$ and convert to memory cells at a rate $k$, which decline at a rate $\beta$. However, before $T_{\max}$, only effector cells grow at a rate $\rho$ and proportionally to a function $F(t)$ which simulates the inclusion with step-wise functions of the co-medication of corticosteroids and tocilizumab (anti–IL-6 receptor antibody). This simple model was able to show the long-term persistence used in CAR-T therapies. 

The authors in \cite{barros2020car} also considered a division between tumour $T$, effector CAR-T cells $C_T$ and memory CAR-T cells $C_M$ in the following model
\begin{subequations}
	\begin{flalign}{}
		\frac{d T}{d t}=& T f(T)-d_{T}\left(T, C_{T}\right), \\
		\frac{d C_{T}}{d t}=& p_{C_{T}}\left(C_{T}\right)-a_{C_{T}}\left(C_{T}\right)+\\
		&+p_{C_{T}}\left(T, C_{M}\right) -d_{C_{T}}\left(T, C_{T}\right),\nonumber \\
		\frac{d C_{M}}{d t}=&p_{C_{M}}\left(C_{M}\right)-d_{C_{M}}\left(T, C_{M}\right)-a_{C_{M}}\left(C_{M}\right),
	\end{flalign}
\end{subequations}
where $f(T)$ is the density dependence growth of tumour cells, and respectively for effector and memory CAR-T cells, we have the following: $p_{C_{T}}(C_T)$ and $p_{C_{M}}(C_M)$ as cell production functions, $d_{C_{M}}\left(T, C_{M}\right)$ and $d_{C_{T}}\left(T, C_{T}\right)$ as cell inhibition functions, and $a_{C_{M}}\left(C_{M}\right)$ and $a_{C_{T}}\left(C_{T}\right)$ as natural death functions. For this model, most functions were considered to be linear, except for the tumour growth function, considered to be logistic growth. Simulations were run for mice data found in the literature, showing different outcomes depending on tumour burden or initial therapy dose. The authors considered that a high CAR-T cell inhibition from tumour leads to tumour escape and absence of CAR-T cell memory. The same CAR-T cell division was considered in the model from \cite{Hanson2016}, not only showing a distinction between effector and memory, but also between the cytotoxic (CD8\textsuperscript{+}) and helper (CD4\textsuperscript{+}) cells. Again, parameter values were not obtained from actual data, but from simulated clinical data. Their results suggest the hypothesis that initial tumour burden is a stronger predictor of toxicity than the initial dose of CAR-T cells. Also,  the authors considered an inflammatory immune response regulated via a Hill function to maintain a realistic bound on the activation rate of T cells. This function gave rise to tumour-burden-correlated toxicity, while the correlation of CAR-T cell dose alone and toxicity was poor.

\begin{figure*}[!hbt]
	\centering
	\includegraphics[width=\textwidth]{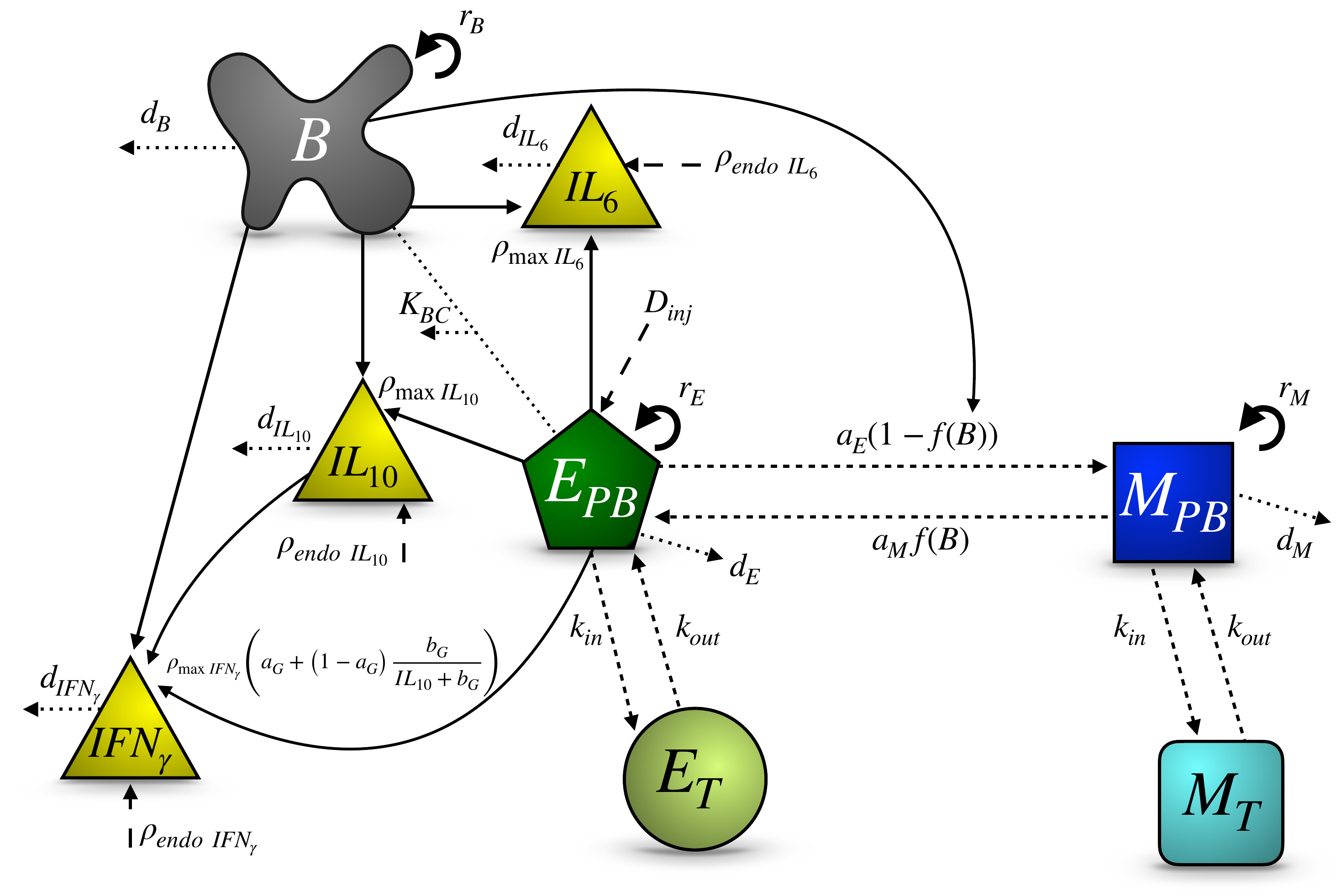}
	\caption{Illustration of the dynamics in model \eqref{model Hardi}. The grey irregular shape represents tumour B-cells in CLL.  Yellow triangles represent the inflammatory cytokines $IL_6=IL_6(t)$, $IL_{10}=IL_{10}(t)$ and interferon $IFN_\gamma=IFN_\gamma(t)$. The green circle and pentagon represent respectively effector CAR-T cells from the PB and from the tissue. Squared, blue shapes represent memory CAR-T cells, also from PB and tissue.  Solid lines represent promotion of cell production, while dotted lines represent cell loss due to natural death or due to encounters between cells. Short-dashed lines represent exchange between cell compartments, and finally long-dashed lines represent constant production in the cell compartments.}
	\label{fig CART}
\end{figure*}

The pharmacological model in \cite{Hardiansyah2019} considered both the influence of CAR-T cells in inflammatory responses with cytokines (such as interleukins $IL_6$, $IL_{10}$ or interferon $IFN_\gamma$), as well as the distinction between CAR-T cells into effector and memory cells. This was also done in order to understand toxicity related to cytokine release syndrome. In the model, the variable $B$ represents CLL tumour B cells in peripheral blood (PB). CAR-T cells in PB are divided into effector $E_{PB}$ and memory $M_{PB}$ cells. This division is also performed for the CAR-T cells in the tissue compartments ($E_T$ and $M_T$). The complete mathematical model is shown in Fig. \ref{fig CART}, and reads

\begin{subequations}
	\label{model Hardi}
	\begin{flalign}{}
		\label{HardiEq1}
		\frac{d B_{P B}}{d t}=&r_{B} B_{P B}-d_{B}  B_{P}-K_{B C}  E_{P B}  B_{P B},\\
		\label{HardiEq2}
		\frac{d IL_{6}}{d t}=&\rho_{\text {endo } \text{IL}_6}+\rho_{\max \text{IL}_{6}}  B_{P B}  E_{P}-d_{\text{IL}_{6}}  I L_{6} ,\\
		\label{HardiEq3}
		\frac{d IL_{10}}{d t}=&\rho_{\text {endo } \text{IL}_{10}}+\rho_{\max \text{IL}_{10}}  B_{P}  E_{P}-d_{\text{IL}_{10}}  \text{IL}_{10},\\
		\label{HardiEq4}
		\frac{d I F N_{\gamma}}{d t}=&\rho_{\text {endo }\text{IFN}_{\gamma}}-d_{\text{IFN}_{\gamma}}  I F N_{\gamma}+\\
		&+\rho_{\max \text{IFN}_{\gamma}}  B_{P B}  E_{P B} \left(a_{G}+\left(1-a_{G}\right)\frac{b_{G}}{I L_{10}+b_{G}}\right)\nonumber,\\
		\label{HardiEq5}
		\frac{d E_{P B}}{d t}=&D_{\text {inj }}+r_{E} E_{P B}  B_{P B}-d_{E}  E_{P B}-k_{i n} E_{P B} +\\
		&+k_{\text {out}} E_{T}-a_{E} E_{P B} (1-f(B_{P B}))+a_{M}  M_{P B}  f(B)\nonumber,\\
		\label{HardiEq6}
		\frac{d E_T}{d t}=&k_\text{in}   E_{P B}-k_\text{out}  E_{T},\\
		\label{HardiEq7}
		\frac{d M_{P B}}{d t}=&r_{M}  M_{P B}-d_{M}  M_{P B}-k_\text{in} M_{P B}+k_\text{out}M_{T} +\\
		&+a_{E}  M_{P B} (1-f(B_{P B}))-a_{M}  M_{P B}  f(B_{P B})\nonumber,\\
		\label{HardiEq8}
		\frac{dM_\text{T}}{d t}=&k_\text{in} M_{P B}-k_\text{out} M_{T}.
	\end{flalign}
\end{subequations}

In this model, parameters $r_B,r_E$ and $r_M$ represent growth rates, while $d_B,d_E$ and $d_M$ are death rate constants, respectively for $B$, $E_{PB}$ and $M_{PB}$ cells. Parameter $K_{BC}$ is the is the effector CAR-T-mediated B-cell CLL degradation rate constant in peripheral blood. For the inflammatory immune responses we have, respectively for $IL_6$, $IL_{10}$ and $IFN_\gamma$ the following constants: $\rho_{\text{endo }\text{IL}_6}, \rho_{\text{endo }\text{IL}_{10}}$ and $\rho_{\text{endo }\text{IFN}_\gamma}$ as endogenous synthesis rates; parameters $\rho_{\text{max }\text{IL}_6}, \rho_{\text{max }\text{IL}_{10}}$ and $\rho_{\text{max }\text{IFN}_\gamma}$ as production rates; and finally,  $d_{\text{IL}_6}, d_{\text{IL}_{10}}$ and $d_{\text{IFN}_\gamma}$ are the natural death rates by the activated CAR-T cells.  Constants $a_G$ and $b_G$ are the inhibitory parameters of $IL_{10}$ on $IFN_\gamma$ production. PB and tissue compartments are distributed via rate constants $k_\text{in}$ and $k_\text{out}$ after intravenous infusion. Peripheral blood effector memory CAR-T cells are activated via activation rates $a_E$ and $a_M$. Finally, function $f(B_{P B})$ is chosen as a Hill function such that $f(B_{P B})=\frac{B_{P B}}{B_{P B}+h},$
with $h$ the half-saturation constant of the tumour. This model was adjusted to data from 3 patients obtained from the literature. Its main conclusion is that toxic inflammatory response is correlated to disease burden, i.e. the number of tumour cells in bone marrow, and not with CAR-T cells doses, contrary to what is observed with most cancer chemotherapies. Other models have also considered these hypotheses, such as the discretised model in \cite{toor2019dynamical}  for CAR-T cells. In this study, a logistic equation of growth was considered to explain the interaction between CAR-T cells and malignant tumour cells. The binding affinity of the CAR-T cell construct (the so-called single-chain variable fragment) and the antigenic epitope (the molecule binding to the antibody) on the malignant target  was considered a critical parameter for all T-cell subtypes modelled.  Both studies show the need for CAR-T cell doses to account for tumour burden, which would require a relatively low number of infused CAR-T cells to achieve the desired target.

\section{Theoretical studies of leukaemia treatment models.}

\label{Section treatment general} 

In previous sections we have described leukaemia growth and response to therapy models that are careful to account for experimental facts or available data. There have been also many studies of models that focus their attention more on methodological mathematical aspects, and provide insight of a more fundamental type. For instance, some of them do not specify which type of leukaemia or treatment they describe.


For instance, some optimal control problems for general leukaemia treatment models have been discussed in the literature.  In \cite{Bratus2012} the authors describe the dynamics of a healthy cell population $N(t)$, a leukaemic population $L(t)$ and a drug $h(t)$ governed by the equations
\begin{subequations}
	\begin{flalign}
		\dfrac{dL(t)}{dt}=&r_l L(t) \ln\left(\dfrac{L_a}{L(t)}\right)-\gamma_l L(t)- f_l(h) L(t),\vspace{5pt}\\
		\dfrac{dN(t)}{dt}=&r_n N(t) \ln\left(\dfrac{N_a}{N(t)}\right)-\gamma_n N(t)-\vspace{5pt}\\
		&-\dfrac{c N(t) L(t)}{1+ L(t)}-f_n(h) N(t),\nonumber\\
		\dfrac{d h(t)}{dt}=&-\gamma_h h(t)+u(t),
	\end{flalign}
\end{subequations}
for $L(0)=L_0,N(0)=N_0, h(0)=0$ and $\gamma_h$ the drug dissipation rate. The effect of the drug was described differently for diseased and healthy cells by the therapy functions $f_l(h)$ and $f_n(h)$, respectively. Here, 
$L_a$ and $N_a$ were the maximum number of diseased and healthy cells respectively, and $\gamma_l$ and $\gamma_n$ were respectively the death rates for the two kinds of cells. Interaction between these subsets was expressed by the parameter $c$. Finally, the control function $u(t)$ is the quantity of drug given to the patient. The authors solved the optimal control problem using the Pontryagin maximum principle. Later research provided additional results along these lines in \cite{Todorov2014}, by using a non-Gompertz interaction term and several phase constraints. Analysis of the switching points was performed, as well as several simulations. Some optimal therapy protocols are shown by introducing a ‘shifting-variable’, which avoids the violation of the normal cell constraint.


Other studies have considered the combined effect of Haematopoietic Inducing Agents (HIA) and Chemotherapeutic Agents (CTA) on stem cells, with the goal of minimising leukopenia \cite {Mouser2014}. Proliferating ($P$) and non-proliferating cells ($N$) were included in the model:
\begin{subequations}
	\begin{flalign}
		\dfrac{dP}{dt}=&-\gamma P +\beta(N)N-\exp(-\gamma t)\beta(N_{\tau})N_\tau+\\
		&+\beta_{HIA}(P)N-\beta_{C}(P)N,\quad \tau<t,\nonumber\vspace{5pt}\\
		\dfrac{dN}{dt}=&-\left[\beta(N)N+\delta N\right]+2\exp(-\gamma t)\beta(N_\tau)N_\tau-\\
		&-\beta_{HIA}(P)N+\beta_C(P)N,\quad \tau<t\nonumber,
	\end{flalign}
\end{subequations}
for $t<\tau$, where $\tau$ was the time for a cell to complete one cycle of proliferation, $\gamma$ the apoptosis rate, and $\delta$ the random cell loss. The expression $N_\tau$ stood for $N(t-\tau)$, introducing a time delay into the equation, and
\addtocounter{equation}{-1}
\begin{subequations}
	\begin{flalign}
		\setcounter{equation}{2}
		\beta(N)&=\beta_0\dfrac{\theta^n}{\theta^n+N^n};\vspace{5pt}\\
		\beta_{HIA}(P)&=\beta_{0,HIA}\dfrac{\theta_1^m}{\theta_1^m+P^m}g_{HIA}(t);\vspace{5pt}\\
		\beta_{C}(P)&=\beta_{0,C}\dfrac{P^w}{\theta_2^w+P^w}g_{C}(t);
	\end{flalign}
\end{subequations}
were Hill functions measuring the rate of cell re-entry into proliferation, the effect of HIA, and the effect of CTA on stem cells, respectively. Also,
\addtocounter{equation}{-1}
\begin{subequations}
	\begin{flalign}
		\setcounter{equation}{5}
		g_{HIA}(t)=\left\{\begin{array}{ll}
			1,&0<t\leq\tau_1,\\
			\exp(-s_1(t-\tau_1)),&t>\tau_1,
		\end{array}\right.
	\end{flalign}
\end{subequations}
simulated the time decay of HIA. Finally, CTA time decay was modelled by
\addtocounter{equation}{-1}
\begin{subequations}
	\begin{flalign}
		\setcounter{equation}{6}
		g_{C}(t)=\left\{\begin{array}{ll}
			1,&0<t\leq\tau_2,\\
			\exp(-s_2(t-\tau_2)),&\tau_2<t\leq\tau_3,\\
			\exp(-s_3(t-\tau_3)),&t>\tau_3.
		\end{array}\right.
	\end{flalign}
\end{subequations}
Using this set of equations the authors found that HIA administration increases the nadir observed in the proliferative cell line compared with when CTA treatment alone is administered. This is significant in preventing patients undergoing chemotherapy treatment from experiencing secondary effects. Furthermore, the steady state value of the proliferating cells was found to be significantly lower in silico after CTA treatment. The model and accompanying analysis give rise to an interesting question: Is concurrent administration of an HIA during chemotherapy a prudent approach for reducing toxicity during chemotherapy? There is substantial clinical evidence to suggest that HIAs could be useful in cases of anemia. They argued that prophylactic benefits of HIAs use together with chemotherapeutic agents at the onset of treatment, although rational, should be balanced with the treatment cost and the risk that HIAs will cause adverse side effects such as venous thromboembolism and tumour progression.

\section{Conclusion.}
Mathematical models have proved to be a essential asset in biomedicine. Haematological diseases are well suited to mathematical modelling, not only with differential equations, but also with stochastic models or other techniques. Therefore, there is a huge amount of data to combine with the mathematical models already in the current literature. Even so, these models may not be sufficient to characterise specific disease behaviours in leukaemia diagnosis: one could take, for example, acute lymphoblastic leukaemia dynamics as a particularly undeveloped issue, as studies of chronic myeloid leukaemia appear to us to have attracted more attention. This is probably because myeloid malignancies are most common in adults. 

Despite the importance of the models presented, the only way to integrate them into clinical practice successfully is to through collaboration between mathematicians, biomedical scientists and clinicians. This can lead to new questions and conclusions for both mathematical models and biological problems. The development of such a useful weapon against cancer should be unified, so that the models can be useful for the actual observation and treatment of disease in patients, beyond the theoretical framework. Mathematical models require refinement in terms of being included in hospital protocols, as a diagnostic or prognostic tool and this can only be achieved by cooperation between the mathematical and medical world.

\begin{acknowledgements}

This work has been partially supported by Ministery of Science and Technology, Spain (grant number PID2019-110895RB-I00), the Spanish Foundation for Science and Technology (FECYT project PR214), the Asociaci\'on Pablo Ugarte (APU, Spain), Junta de Andaluc\'ia (Spain) group FQM-201, Junta de Comunidades de Castilla-La Mancha (grant number SBPLY/17/180501/000154), the Programme of Research and Transfer Promotion from the University of C\'adiz (grant number EST2020-025) and Health and Family department of Junta de Andaluc\'{\i}a/FEDER (grant number ITI-0038-2019).

\end{acknowledgements}

%
 \section*{Conflict of interest}

 The authors declare that they have no conflict of interest.

\bibliographystyle{spmpsci}      
\bibliography{general.bib}

\end{document}